\documentclass[preprint2]{aastex62}

\received{}
\revised{}
\accepted{}
\submitjournal{}

\shorttitle{Synthetic Observations of \ion{H}{1} Line Profiles}
\shortauthors{Fukui et al.}

\begin{document}

\title{Synthetic Observations of 21\,cm \ion{H}{1} Line Profiles from Inhomogeneous Turbulent Interstellar \ion{H}{1} Gas with Magnetic Field}

\correspondingauthor{Yasuo Fukui}
\email{fukui@a.phys.nagoya-u.ac.jp}

\author{Yasuo Fukui}
\affiliation{Department of Physics, Nagoya University, Furo-cho, Chikusa-ku, Nagoya 464-8602, Japan}

\author{Takahiro Hayakawa}
\affiliation{Department of Physics, Nagoya University, Furo-cho, Chikusa-ku, Nagoya 464-8602, Japan}
\affiliation{Department of Physical Science, Osaka Prefecture University, 1-1 Gakuen, Sakai, Osaka 599-8531, Japan}

\author{Tsuyoshi Inoue}
\affiliation{Department of Physics, Nagoya University, Furo-cho, Chikusa-ku, Nagoya 464-8602, Japan}

\author{Kazufumi Torii}
\affiliation{Nobeyama Radio Observatory, National Astronomical Observatory of Japan, 462-2 Nobeyama, Minamimaki, Minamisaku, Nagano 384-1305, Japan}

\author{Ryuji Okamoto}
\affiliation{Department of Physics, Nagoya University, Furo-cho, Chikusa-ku, Nagoya 464-8602, Japan}

\author{Kengo Tachihara}
\affiliation{Department of Physics, Nagoya University, Furo-cho, Chikusa-ku, Nagoya 464-8602, Japan}

\author{Toshikazu Onishi}
\affiliation{Department of Physical Science, Osaka Prefecture University, 1-1 Gakuen, Sakai, Osaka 599-8531, Japan}

\author{Katsuhiro Hayashi}
\affiliation{Department of Physics, Nagoya University, Furo-cho, Chikusa-ku, Nagoya 464-8602, Japan}



\begin{abstract}

We carried out synthetic observations of interstellar atomic hydrogen at 21\,cm wavelength by utilizing the magneto-hydrodynamical numerical simulations of the inhomogeneous turbulent interstellar medium (ISM) \citep{2012ApJ...759...35I}.
The cold neutral medium (CNM) shows significantly clumpy distribution with a small volume filling factor of 3.5\%, whereas the warm neutral medium (WNM) distinctly different smooth distribution with a large filling factor of 96.5\%.
In projection on the sky, the CNM exhibits highly filamentary distribution with a sub-pc width, whereas the WNM shows smooth extended distribution.
In the \ion{H}{1} optical depth the CNM is dominant and the contribution of the WNM is negligibly small.
The CNM has an area covering factor of 30\% in projection, while the WNM has a covering factor of 70\%.
This causes that the emission-absorption measurements toward radio continuum compact sources tend to sample the WNM with a probability of 70\%, yielding smaller \ion{H}{1} optical depth and smaller \ion{H}{1} column density than those of the bulk \ion{H}{1} gas.
The emission-absorption measurements, which are significantly affected by the small-scale large fluctuations of the CNM properties, are not suitable to characterize the bulk \ion{H}{1} gas.
Larger-beam emission measurements which are able to fully sample the \ion{H}{1} gas will provide a better tool for that purpose, if a reliable proxy for hydrogen column density, possibly dust optical depth and gamma rays, is available.
The present results provide a step toward precise measurements of the interstellar hydrogen with $\sim 10$\% accuracy.
This will be crucial in the interstellar physics including identification of the proton-proton interaction in gamma-ray supernova remnants.
\end{abstract}




\section{INTRODUCTION}\label{sec:introduction}
The main constituent of the interstellar medium (ISM) is atomic hydrogen \ion{H}{1}, and the secondary constituents, whose abundance is ten times less than \ion{H}{1}, include molecular hydrogen H$_{2}$ and atomic helium He over the global volume of the Galactic disk.
It is of primary importance to make precise measurement of \ion{H}{1} in our understanding of the structure, kinematics and physical conditions of the interstellar medium and the formation of interstellar clouds and stars.

The 21\,cm spin flip transition of \ion{H}{1} offers a direct method to measure interstellar \ion{H}{1} and has been used extensively over the last several decades since its discovery in 1951 \citep{1951Natur.168..356E,1951Natur.168..357M}.
When the \ion{H}{1} 21\,cm line is optically thin, the following equation is used to calculate the \ion{H}{1} column density, $N_{\text{H$\;$\sc{i}}}$, from the 21\,cm line intensity, $W_{\text{H$\;$\sc{i}}}$,
\begin{equation}\label{eqn:opticallythin_eqn}
N_{\text{H$\;$\sc{i}}}\,\text{(cm$^{-2}$)}=1.823\times 10^{18}\,W_{\text{H$\;$\sc{i}}}\,\text{(K\,km\,s$^{-1}$)}.
\end{equation}
As such, it has been commonly thought that 21\,cm \ion{H}{1} emission is optically thin.
Direct support for the optically thin assumption for \ion{H}{1} is obtained by the emission-absorption measurements toward radio continuum compact sources, which show that the \ion{H}{1} peak optical depth is typically $\sim$0.1 \citep[e.g.,][]{2003ApJ...585..801D,2003ApJS..145..329H,2003ApJ...586.1067H}.
High resolution \ion{H}{1} observations with the Arecibo 305\,m telescope have been used to make high sensitivity emission-absorption measurements and revolutionized the knowledge on the \ion{H}{1} gas physical conditions \citep{2003ApJ...586.1067H}.
In the meantime the question was raised that the 21\,cm \ion{H}{1} emission may be optically thick based on \ion{H}{1} profiles with self-absorption \citep{2012ApJ...749...87B}.
Because \ion{H}{1} observations provide physical quantities averaged along a line of sight, it is in principle impossible to retrieve the original physical parameters of the \ion{H}{1} gas in the three dimensions, making it difficult to test observationally the above \ion{H}{1} properties for the large volume where \ion{H}{1} is distributed.

The dust emission and extinction are also used often as a proxy for $N_{\text{H$\;$\sc{i}}}$ under an assumption of constant gas to dust ratio.
Previously, the scattering in the data for dust column density against $W_{\text{H$\;$\sc{i}}}$ was large, making the method crude at best \citep[see e.g., Chapter 21 of][]{2011piim.book.....D}.
\citet{2014A+A...571A..11P} opened a new possibility of precise measurement of dust optical depth by making extremely sensitive measurements of dust optical depth at sub-mm wavelengths, 350, 550 and 850\,microns.
These long wavelengths are in the Rayleigh-Jeans regime of the Planck function and, by combining with the \textit{IRAS} data at 100 microns in the Wien regime, the sub-mm dust optical depth and dust temperature for an appropriate dust emissivity $\beta$ were calculated with unprecedented accuracy to within 10\%.

\citet{2014ApJ...796...59F,2015ApJ...798....6F} presented a method to use the \textit{Planck}/\textit{IRAS} dust optical depth at 353\,GHz ($\tau_{353}$) as a proxy of $N_{\text{H$\;$\sc{i}}}$ by identifying the optically thin regime of 21\,cm \ion{H}{1} emission as a linear part of a scatter plot between $W_{\text{H$\;$\sc{i}}}$ and $\tau_{353}$, where dispersion of the data points is smallest at the highest dust temperature.
\citet{2014ApJ...796...59F} presented results for high-latitude clouds with the Galactic Arecibo L-band Feed Array \ion{H}{1} (GALFA-\ion{H}{1}) survey data \citep{2011ApJS..194...20P} taken with a $4\arcmin$ beam of the Arecibo telescope and \citet{2015ApJ...798....6F} for the whole sky at $|b|$ larger than $15\arcdeg$ with a $33\arcmin$ beam in the Leiden/Argentine/Bonn (LAB) survey \citep{2005A&A...440..775K}.
The two papers concluded that, in the local interstellar volume within 200\,pc of the sun, interstellar \ion{H}{1} is dominated by cold and dense \ion{H}{1} gas which is optically thick with a typical \ion{H}{1} optical depth of $\sim$1, and that the average \ion{H}{1} density is to be doubled approximately if the correction for the optical depth is applied.
\citet{2015ApJ...798....6F} argued that the opacity-corrected \ion{H}{1} can explain the ``dark gas'', which is detected in $\gamma$-rays and interstellar extinction $A_{V}$ but not in the 2.6-mm CO or optically-thin 21-cm \ion{H}{1} transitions (\citealt{2005Sci...307.1292G}; \citealt{2015ARA&A..53..199G} for a review), as an alternative to CO-free H$_2$ gas \citep{2010ApJ...716.1191W}.
In order to understand the behavior of \ion{H}{1}, it is crucial to measure the fraction of H$_2$ in \ion{H}{1} gas.
Since H$_2$ has no radio transition, ultraviolet (UV) absorption of the electronic transition provides a unique tool to directly measure H$_2$.
\textit{FUSE} and \textit{Copernicus} results are such datasets of H$_2$ \citep[e.g.,][]{2006ApJ...636..891G}.
Since observations need background UV sources which are located at high $b$, the H$_2$ observations measure H$_2$ in the local interstellar medium close to the sun.
We are able to use the H$_2$ data in modeling the local interstellar medium.
In some cases \ion{H}{1} can be measured as well in UV. 
Also, \ion{H}{1} measurements at 21\,cm in line absorption toward radio continuum sources provide \ion{H}{1} column density \citep[e.g.,][]{2003ApJS..145..329H,2003ApJ...586.1067H}.

There remain two issues which were not addressed in \citet{2014ApJ...796...59F,2015ApJ...798....6F}.
One is the contribution of the warm neutral medium (WNM).
Pioneering studies by \citet{1965ApJ...142..531F} and \citet{1969ApJ...155L.149F} showed that the interstellar medium consists of the two phases, the CNM and the WNM, which are in pressure equilibrium.
Because the dust grains are included in the both phases, the cold neutral medium (CNM) and WNM, the \ion{H}{1} emission analyzed with the \textit{Planck}/\textit{IRAS} data should include the contribution of the WNM.
The other is the possible effect of dust evolution found by \citet{2013ApJ...763...55R} which may require some modification of the linear relationship between $N_{\text{H$\;$\sc{i}}}$ and $\tau_{353}$ assumed by \citet{2014ApJ...796...59F,2015ApJ...798....6F}.
\ion{H}{1} emission-absorption measurements were used to constrain \ion{H}{1} parameters of the CNM and WNM \citep{2003ApJ...585..801D,2003ApJS..145..329H,2003ApJ...586.1067H}, where the WNM manifests itself as broad line wings of \ion{H}{1} emission profiles.
There remains yet an uncertainty in deriving the WNM temperature in absorption, and only a lower limit for the spin temperature $T_\mathrm{s}$ was obtained to be around 500\,K, leaving the mass of the WNM uncertain, which may occupy $\sim 60$\% of total \ion{H}{1} \citep{2003ApJ...586.1067H}. In addition, the spatial distribution of the CNM and the WNM is not clearly understood yet while the CNM is suggested to occupy smaller volume than the WNM \citep{1965ApJ...142..568F}.

Following \citet{2014ApJ...796...59F,2015ApJ...798....6F}, \citet{2014ApJ...793..132S} made \ion{H}{1} emission-absorption measurements toward radio continuum sources in Perseus with the Arecibo \ion{H}{1} data and found that the absorption optical depth is not so large as suggested by \citet{2014ApJ...796...59F,2015ApJ...798....6F}, raising a question on the optically thick \ion{H}{1} emission.
Their results are consistent with those by \citet{2003ApJS..145..329H,2003ApJ...586.1067H}.
\citet{2015ApJ...814...13M} made a comparison of \citet{2015ApJ...798....6F} with the \ion{H}{1} model by \citet{1981ApJ...247L..73H} and discussed that the two results are consistent within $\sim$10 \% in spite of their different \ion{H}{1} optical depth.
The reason for this agreement is not clarified.
The method by \citet{2015ApJ...798....6F} is based on a simple assumption of uniform interstellar medium and may need modification if realistic non-uniform physical properties of the interstellar medium are taken into account.
The real \ion{H}{1} observations are, however, limited because we are not able to assess the actual three dimensional physical conditions of the \ion{H}{1} gas emitting/absorbing 21\,cm line radiation. 

A possible solution to overcome the difficulty and to test the above discrepancy is to utilize the results of hydrodynamical numerical simulations of the \ion{H}{1} gas \citep{2015ApJ...804...89M,2016arXiv161202017M}.
Recently, three-dimensional hydrodynamical simulations modeled converging \ion{H}{1} flows and achieved realistic density distributions and kinematics with high inhomogeneity and strong turbulence \citep{2008A&A...486L..43H,2009ApJ...695..248H,2009MNRAS.398.1082B,2011MNRAS.414.2511V,2012ApJ...759...35I,2014ApJ...786...64K}.
These simulations are supported by observations of nearby galaxies which show turbulent \ion{H}{1} gas with density of 10--100\,cm$^{-3}$ and molecular clouds formed from \ion{H}{1} gas \citep{2006ApJ...650..933B,1999PASJ...51..745F,2008ApJS..178...56F,2009ApJ...705..144F,2009ApJS..184....1K,2010ARA&A..48..547F}.

In order to clarify the cause of the difference between the emission-absorption measurements of \ion{H}{1} and the \textit{Planck}/\textit{IRAS}-based method of \citet{2014ApJ...796...59F,2015ApJ...798....6F} and to have a better understanding of the CNM and the WNM, we examine synthetic \ion{H}{1} line profiles by using the data of magneto-hydrodynamical (MHD) numerical simulations where the density, temperature, and velocity of the \ion{H}{1} gas are available in three dimensions \citep{2012ApJ...759...35I}.
These simulations deal with converging \ion{H}{1} flows as a function of time over 10\,Myrs.
The gas is originally \ion{H}{1}, while formation of H$_2$ molecules is incorporated by using the usual dust surface reaction.
The results indicate two phases of \ion{H}{1}, the CNM and the WNM, as well as time-dependent transient gas which behaves intermediately.
In the following we call for convenience the gas with $T_\mathrm{s}$ below 300\,K the CNM and that with $T_\mathrm{s}$ above 300\,K the WNM.

In the present paper we focus on the spatial distribution of the \ion{H}{1} gas derived from the synthetic observations and explore the astrophysical implications of the emission-absorption measurements on the \ion{H}{1} properties.
Another paper which compares the synthetic observations and \citet{2014ApJ...796...59F,2015ApJ...798....6F} will be published separately. 
The paper is organized as follows; Section \ref{sec:simulationmodel} gives the results of the simulations, Section \ref{sec:observedprops} presents results of synthetic observations with discussion and Section \ref{sec:discussion} describes the spatial distribution of the \ion{H}{1} optical depth and column density with discussion.
In Section \ref{sec:conclusions} we present the conclusions.

\begin{table*}
\begin{center}
\caption{Summary of symbols in the text}\label{tab:para_text}
\begin{tabular}{ll}
\tableline\tableline
Symbol & Description \\
\tableline
$n_{X}$ & Number densities of atomic/molecular species, $X=$\ion{H}{1}, H$_2$ etc. \\
$\tau_{353}$ & Dust optical depth at 353\,GHz by \citet{2014A+A...571A..11P} \\
$T_\mathrm{k}$ & Kinetic temperature of gas \\
$\tau_{\text{H$\;$\sc{i}}}$, $T_\mathrm{s}$ & \ion{H}{1} optical depth and spin temperature \\
$\tau_{\text{H$\;$\sc{i}}}^\mathrm{model}$ & Model \ion{H}{1} optical depth given by Equation (\ref{eqn:real_tau}) \\
$\tau_{\text{H$\;$\sc{i}}}^\mathrm{abs}$ & \ion{H}{1} optical depth obtained by emission-absorption measurements (Eq.\ \ref{eqn:exptauabs}) \\
$\langle T_\mathrm{s}\rangle$ & Density-weighted harmonic mean of $T_\mathrm{s}$ along a line-of-sight (Eq.\ \ref{eqn:harmonicmean_ts})\\
$T_\mathrm{b}$ & Brightness temperature of \ion{H}{1} spectrum \\
$W_{\text{H$\;$\sc{i}}}$ & Velocity integrated-intensity of \ion{H}{1} spectrum \\
$W_{\text{H$\;$\sc{i}}}^\mathrm{WNM}$ & WNM integrated-intensity \\
 & calculated by setting $\epsilon=0$ of the CNM, while $\kappa$ is held fixed \\
$W_{\text{H$\;$\sc{i}}}^\mathrm{CNM}$ & CNM integrated-intensity defined by $W_{\text{H$\;$\sc{i}}}^\mathrm{CNM}=W_{\text{H$\;$\sc{i}}}-W_{\text{H$\;$\sc{i}}}^\mathrm{WNM}$ \\
$N_{\text{H$\;$\sc{i}}}$ & \ion{H}{1} column density \\
$N_{\text{H$\;$\sc{i}}}^\mathrm{model}$ & Model \ion{H}{1} column density given by $\sum_{j} n_{\text{H$\;$\sc{i}}, j}\Delta y$ \\
$N_{\text{H$\;$\sc{i}}}^\mathrm{thin}$ & \ion{H}{1} column density obtained under assumption of optically-thin \ion{H}{1} line (Eq.\ \ref{eqn:opticallythin_eqn}) \\
$N_{\text{H$\;$\sc{i}}}^\mathrm{HT}$ & \citet{2003ApJS..145..329H} \ion{H}{1} column density \\
$f_\mathrm{H2}$ & Molecular fraction defined as $f_\mathrm{H2}=2N_\mathrm{H2}/(2N_\mathrm{H2}+N_{\text{H$\;$\sc{i}}})$ or $f_\mathrm{H2}=2n_\mathrm{H2}/(2n_\mathrm{H2}+n_{\text{H$\;$\sc{i}}})$ \\
$M_{\text{H$\;$\sc{i}}}$ & Mass of \ion{H}{1} \\
$M_{\text{H$\;$\sc{i}}}^\mathrm{thin}$ & Mass of \ion{H}{1} obtained from $N_{\text{H$\;$\sc{i}}}^\mathrm{thin}$ \\
\tableline
\end{tabular}
\end{center}
\end{table*}

\begin{table*}
\begin{center}
\caption{Summary of the physical parameters in the MHD model}\label{tab:para_model}
\begin{tabular}{ll}
\tableline\tableline
Symbol & Description \\
\tableline
$pk_\mathrm{B}^{-1}$ & Thermal pressure of gas, $T_\mathrm{k}=pk_\mathrm{B}^{-1}\left(\sum_{X}n_{X}\right)^{-1}$ \\
$(V_x, V_y, V_z)$ & Velocity vector \\
\tableline
\end{tabular}
\end{center}
\end{table*}

\section{RESULTS OF SIMULATIONS}\label{sec:simulationmodel}

\subsection{Simulation Data and Model Selection}
\begin{figure}
\includegraphics{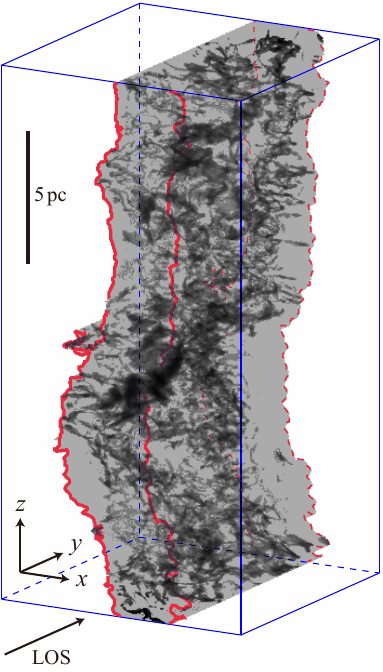}
\caption{
Volume rendering map of spatial distribution of pixels with $T_\mathrm{s} < 300$\,K (black) and those with $T_\mathrm{s} > 300$\,K (gray) in the 0.5-Myr model.
The red contour projected on the front surface outlines the region of interest (ROI).
The pixels out of the ROI are blanked and not used in Figures \ref{fig:NHI-fH2}, \ref{fig:histo_NHI}(a), \ref{fig:histo_n_T}, \ref{fig:histo_W_N_tau}, \ref{fig:NHI-WHI}, \ref{fig:pix-4arcmin}, \ref{fig:model_thin} and Table \ref{tab:mass}.
\label{fig:v-rendering}}
\end{figure}

We summarize the relevant physical parameters and symbols in Table \ref{tab:para_text}.
We then give a brief explanation of the physical parameters and settings of the MHD simulations.
More details are found in \citet{2012ApJ...759...35I}.
The simulations assume converging \ion{H}{1} gas flows at 20\,km\,s$^{-1}$ which are initially in pressure equilibrium with the standard interstellar \ion{H}{1} having pressure of $pk_\mathrm{B}^{-1}=5.2\times 10^3$\,K\,cm$^{-3}$.
The $x$-, $y$- and $z$-axes are taken as in Figure 2 of \citet{2012ApJ...759...35I} and the flow direction is parallel to the $x$-axis.
The \ion{H}{1} gas flow is inhomogeneous and continuously enters into the box from the two opposite boundaries of a cube of (20\,pc)$^{3}$.
In the interface of the converging flows turbulence is excited and the magnetic field is amplified.
Formation of molecules such as H$_2$ formation on dust surfaces and CO formation via CH$_{2}^{+}$ with the effects of self/dust UV shielding are taken into account and radiative and collisional heating and atomic and molecular cooling are incorporated.
The simulation data are provided as the three dimensional data cube with $512^{3}$ uniform pixels\footnote{\citet{2012ApJ...759...35I} made the simulations with dividing the numerical domain into $1024^3$ pixels but the data were provided at a factor-of-two lower resolution to reduce the data size.} and each pixel having a size of 0.04\,pc in each axis with the physical parameters as listed in Table \ref{tab:para_model}.
The simulations are made over a timescale of 10\,Myrs, ten times the typical crossing timescale of the local \ion{H}{1} gas in the solar neighborhood.
The total gas mass in the numerical domain increases with time.
\added{
In order to extract the colliding gas for the present analysis, we excluded the gas injected prior to the collision in the simulation box.
The spatial distribution of the colliding gas is localized around the central part of the box in the $x$-axis and is significantly different in its intensity from the injected gas.
In order to eliminate the injected gas we set a lower limit of the synthetic \ion{H}{1} intensity (see section \ref{ssec:syntheticobservations}) at 150\,K\,km\,s$^{-1}$ in the projected $x$-$z$ plane.
Figure \ref{fig:v-rendering} illustrates the boundary determined in that way by red contours in the three dimensional view at an epoch of 0.5\,Myrs.
}

\begin{figure*}
\includegraphics{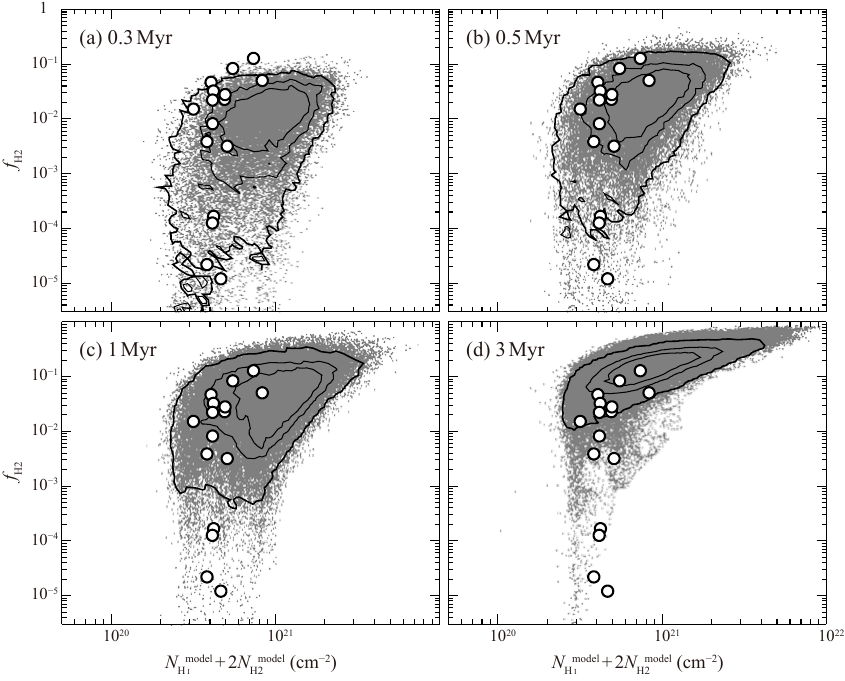}
\caption{
(a) Plot of molecular fraction defined as $f_\mathrm{H2}=2N_\mathrm{H2}/(2N_\mathrm{H2}+N_{\text{H$\;$\sc{i}}})$ for total column density $N_{\text{H$\;$\sc{i}}}+2N_\mathrm{H2}$ at a time step of 0.3\,Myr.
The contours includes 45\%, 70\% and 95\% of data points.
The open circles show the results of direct UV absorption measurements of H$_2$ by \textit{FUSE} toward AGNi (summarized in Table \ref{tab:fH2estimates}, 3 out of 19 are not shown due to low $f_\mathrm{H2}\sim10^{-6}$).
(b)--(d) Same as (a) but at time steps of 0.5\,Myr, 1\,Myr, and 3\,Myr, respectively.
\label{fig:NHI-fH2}}
\end{figure*}

\begin{figure}
\includegraphics{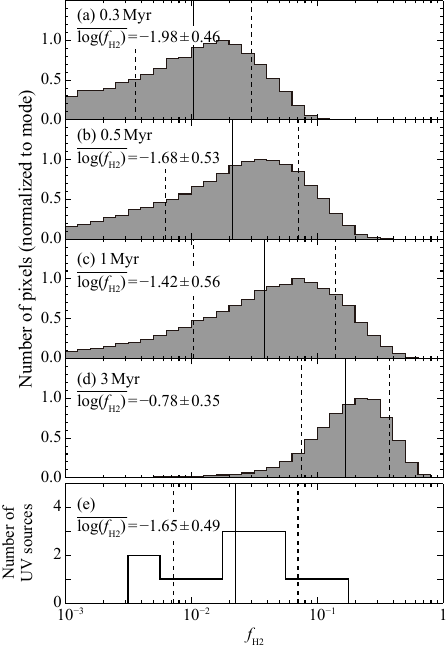}
\caption{
Histograms of $f_\mathrm{H2}$ in the models at four epochs at 0.3\,Myrs, 0.5\,Myrs, 1\,Myrs and 3\,Myrs.
Panel (a) presents the 0.3-Myr model.
The average and dispersion ($1\sigma$) of $f_\mathrm{H2}$ are shown as corresponding bars with of a solid line and two dashed lines, respectively.
Panels (b)--(d) show similar plots at the other epochs.
Panel (e) shows a histogram of $f_\mathrm{H2}$ by the UV measurements listed in Table \ref{tab:fH2estimates}, and presents the same quantities as above.
\label{fig:histo_fH2}}
\end{figure}

\begin{figure}
\includegraphics{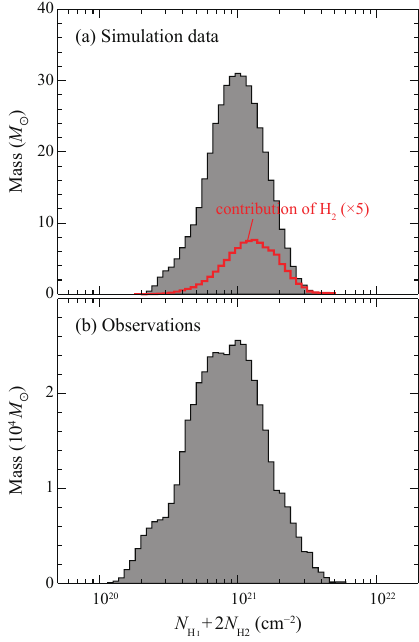}
\caption{
(a) Mass histograms of total column density, $N_{\text{H$\;$\sc{i}}}+2N_\mathrm{H2}$ in the 0.5-Myr model.
The red line represents the contribution of H$_2$ (multiplied by a factor of 5).
(b) Same as (a) but for the observational dataset used in \citet{2015ApJ...798....6F}.
Here the total column densities are given from $\tau_{353}$ \citep{2014A+A...571A..11P} by taking into account a nonlinear relationship.
\label{fig:histo_NHI}}
\end{figure}

Figures \ref{fig:NHI-fH2}(a)--(d) show the distributions of the synthetic data points in the $(N_{\text{H$\;$\sc{i}}}+2N_\mathrm{H2})$-$f_\mathrm{H2}$ plane, where the integration was made for 10\,pc along the $y$-axis.
Data at the four time steps, 0.3, 0.5, 1 and 3\,Myrs, are shown (see the physical parameters in Table \ref{tab:mass}).
UV observations of $f_\mathrm{H2}$ toward extra-galactic sources \citep{2006ApJ...636..891G} are shown by open circles in each panel and are summarized in Table \ref{tab:fH2estimates}, where the number of observed sources for $f_\mathrm{H2}$ is limited to 19.
We did not include Galactic OB stars which may be contaminated by localized gas \citep{2002ApJ...577..221R}, possibly causing unreliable $f_\mathrm{H2}$ values for the local ISM.
The ranges of $N_{\text{H$\;$\sc{i}}}$ and $f_\mathrm{H2}$ are consistent with those of the synthetic data points, whereas the UV measurements are limited to $N_\mathrm{H2}< 10^{21}$\,cm$^{-2}$.
\replaced{Among the four time steps, we find the 0.5-Myr model shows the best presentation of the observations since the fraction of the data points included within a 95\% contour is the largest (14 out of 19).}{
In order to choose the epoch of the model, we compared the measured distribution of $f_\mathrm{H2}$ with the model.
Figures \ref{fig:histo_fH2}(a)--(d) show the corresponding histograms of $f_\mathrm{H2}$ in the model.
Figure \ref{fig:histo_NHI} shows a typical hydrogen column density distribution.
The points of the UV measurements have total hydrogen column density less than $10^{21}$\,cm$^{-2}$, and the comparison was made substantially at $N_{\text{H$\;$\sc{i}}}+2N_\mathrm{H2}$ less than $10^{21}$\,cm$^{-2}$.
Figure \ref{fig:histo_fH2} shows that $f_\mathrm{H2}$ increases by the H$_{2}$ formation reaction in time, while the $N_{\text{H$\;$\sc{i}}}+2N_\mathrm{H2}$ distribution does not change significantly among the epochs (Figure \ref{fig:NHI-fH2}).
By comparing Figures \ref{fig:histo_fH2}(a)-(d) with the histogram of the measured $f_\mathrm{H2}$ in Figure \ref{fig:histo_fH2}(e), we found that the 0.5-Myr model shows the best presentation of the UV measurements in the average and dispersion of $f_\mathrm{H2}$. 
}
We shall use the 0.5-Myr model for the present analysis.
Figure \ref{fig:histo_NHI} gives histograms of total column density $N_{\text{H$\;$\sc{i}}}+2N_\mathrm{H2}$ in the model and $N_{\text{H$\;$\sc{i}}}$ of the observations by \citet{2015ApJ...798....6F}\footnote{\citet{2015ApJ...798....6F} obtained \ion{H}{1} column densities assuming a linear relationship $N_{\text{H$\;$\sc{i}}}\propto \tau_{353}$ but we obtained total column densities by applying a nonlinear relationship $N_{\text{H$\;$\sc{i}}}+2N_\mathrm{H2}\propto \tau_{353}^{1/1.3}$ \citep{2017ApJ...838..132O}.}.
As seen in Table \ref{tab:mass} the fraction of H$_2$ is negligibly small in the 0.5-Myr model.

\subsection{\ion{H}{1} distributions}
\begin{figure*}
\includegraphics{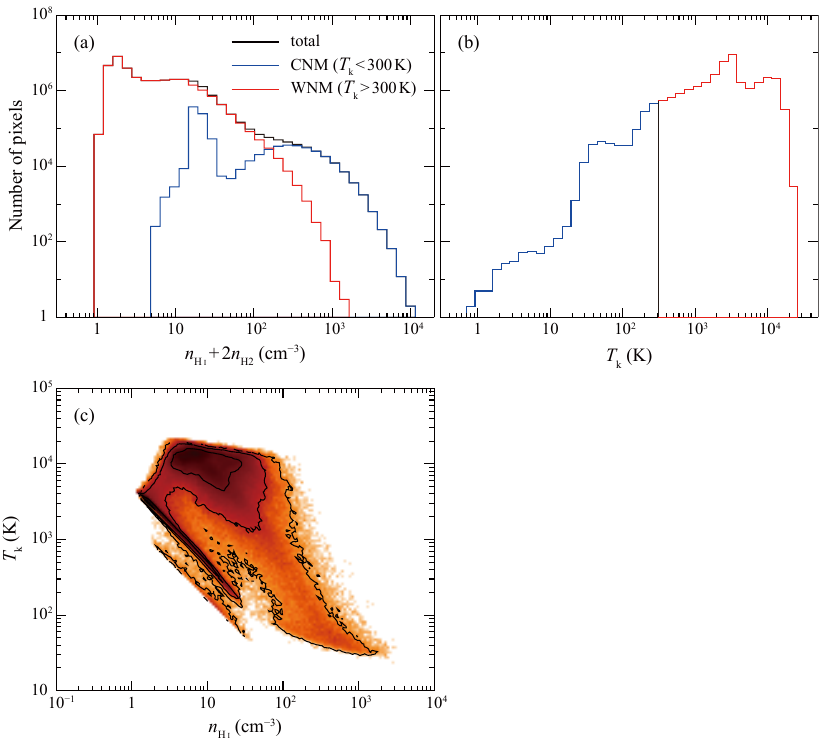}
\caption{
Histograms of (a) total hydrogen density ($n_{\text{H$\;$\sc{i}}}+2n_\mathrm{H2}$) and (b) kinetic temperature ($T_\mathrm{k}$) for each pixel in the 0.5-Myr model.
The blue lines represent the contribution of $T_\mathrm{K}<300$\,K and red lines represent that of $T_\mathrm{K}>300$\,K.
(c) Probability distribution function in the $n_{\text{H$\;$\sc{i}}}$-$T_\mathrm{k}$ plane.
The contours have equal logarithmic spacing.
The straight distribution in the lower right shows the initial \ion{H}{1} flows prior to the collision.
\label{fig:histo_n_T}}
\end{figure*}

Figures \ref{fig:histo_n_T}(a) and \ref{fig:histo_n_T}(b) shows histograms of density ($n=n_{\text{H$\;$\sc{i}}}+2n_\mathrm{H2}$) and temperature ($T_\mathrm{k}$) in the model.
Figure \ref{fig:histo_n_T}(c) shows a probability distribution function in the density-temperature plane and indicates that temperature is roughly inversely proportional to density.
We find that density and temperature have large ranges covering the CNM and WNM, whereas we do not see clear bimodal distribution corresponding to each of the CNM and WNM with a boundary at 300K in Figure \ref{fig:histo_n_T} due to intermediate gas formed by the strong turbulent mixing in the model.
Note that the typical ISM is affected by supernovae with every a few million years and duration of compression (or lifetime of a supernova shock) is about 1 million years.
The ISM compressed by the converging flows in the 0.5--1.0 Myr seems to be the representative state of the dynamic ISM.

Three-dimensional distribution of the model is shown in Figure \ref{fig:v-rendering}.
The CNM and the WNM have distinctly different spatial distributions.
The CNM is highly clumpy with size scales of a few pc to sub-pc, whereas the WNM is diffuse and smooth.
The volumes of the CNM and the WNM are 3.5\% and 96.5\%, respectively.
The CNM is dense gas which quickly cools down, and the WNM has high temperature and high pressure.
These physical properties produce the different spatial distributions.
It is not appropriate to characterize the two media by representative density or temperature because they range over two to three orders of magnitude as seen in Figure \ref{fig:histo_n_T}.
$T_\mathrm{k}=300$\,K as a boundary between the CNM and WNM is consistent with a typical $T_\mathrm{k}\sim 70$\,K of the CNM and a typical $T_\mathrm{k}$ range of the WNM 500\,K--5000\,K \citep{2003ApJ...586.1067H}.
The masses of the CNM and WNM are 150\,$M_\sun$ and 244\,$M_\sun$, respectively in the present model.
The CNM and WNM are far from the dynamical equilibrium which was discussed in a classical picture of the ISM \citep{1965ApJ...142..531F}, but are highly transient and time-dependent (cf.\ Figure \ref{fig:histo_n_T}(c)). 
The time scale of the ISM evolution is in the order of Myr as estimated by a ratio of 10\,km\,s$^{-1}$ divided by 10\,pc.
This is comparable to that of the shock front passage driven by SNRs and is supposed to be usual as the ISM in the solar neighborhood \citep{2012ApJ...759...35I}.

\begin{deluxetable*}{crrrrRRRRR}
\tablewidth{0pc}
\tablecaption{Physical parameters from the models at different time steps\label{tab:mass}}
\tablehead{
\colhead{Time step} & \colhead{$M_{\text{H$\;$\sc{i}}}$} & \colhead{$M_{\text{H$\;$\sc{i}}}^\mathrm{thin}$} & \colhead{$\frac{M_{\text{H$\;$\sc{i}}}}{M_{\text{H$\;$\sc{i}}}^\mathrm{thin}}$} & \colhead{$M_\mathrm{H2}$} & \colhead{$\frac{M_\mathrm{H2}}{M_\mathrm{H2}+M_{\text{H$\;$\sc{i}}}}$} & \colhead{$\frac{M_\mathrm{CNM}}{M_{\text{H$\;$\sc{i}}}}$} & \colhead{$\frac{M_\mathrm{WNM}}{M_{\text{H$\;$\sc{i}}}}$} & \colhead{CNM volume} & \colhead{CNM covering} \\
\colhead{(Myr)} & \colhead{($M_\sun$)} & \colhead{($M_\sun$)} & & \colhead{($M_\sun$)} & & & & \colhead{filling factor} & \colhead{factor}
}
\decimalcolnumbers
\startdata
0.3 & 265 & 205 & 1.3 & 4.5 & 1.7\times 10^{-2}  & 0.39 & 0.61 & 3.4\% & 30.4\% \\
0.5 & 394 & 309 & 1.3 & 18.8 & 4.6\times 10^{-2} & 0.38 & 0.62 & 3.5\% & 28.5\% \\
1.0 & 745 & 563 & 1.3 & 74.6 & 9.1\times 10^{-2} & 0.39 & 0.61 & 5.0\% & 35.7\% \\
3.0 & 1731 & 1114 & 1.6 & 660.5 & 2.8\times 10^{-1} & 0.60 & 0.40 & 12.6\% & 50.6\% \\
\enddata
\tablecomments{
Columns (2): mass of \ion{H}{1} gas, (3): mass of \ion{H}{1} given from \ion{H}{1} integrated-intensity under the optically-thin assumption , (4): ratio of (2) to (3), (5): mass of H$_2$ gas, (6): mass fraction of H$_2$ gas, (7): ratio of CNM mass to (2), (8): ratio of WNM mass to (2), (9): volume filling factor of the CNM, (10): projected area covering factor of the CNM with $\int \tau_{\text{H$\;$\sc{i}}}^\mathrm{model}(V)dV >4$\,km\,s$^{-1}$ (see sections \ref{sec:observedprops} and \ref{ssec:cnmfilaments}).}
\end{deluxetable*}

\begin{deluxetable*}{lRRRRRR}
\tablecolumns{7}
\tablewidth{0pc}
\tablecaption{Targets of the $f_\mathrm{H2}$ estimates\label{tab:fH2estimates}}
\tablehead{
\colhead{Target} & \colhead{$l$} & \colhead{$b$} & \colhead{$N_\mathrm{H2}$} & \colhead{$\tau_{353}$} & \colhead{$N_{\text{H$\;$\sc{i}}}+2N_\mathrm{H2}$} & \colhead{$f_\mathrm{H2}$} \\
 & & & \colhead{(cm$^{-2}$)} & & \colhead{(cm$^{-2}$)} }
\colnumbers
\startdata
3C 249.1 & 130\fdg 39 & +38\fdg 55 & 9.5\times 10^{18} & 2.51\times 10^{-6} & 4.1\times 10^{20} & 4.7\times 10^{-2} \\
ESO 141$-$G55 & 338\fdg 18 & -26\fdg 71 & 2.1\times 10^{19} & 6.38\times 10^{-6} & 8.3\times 10^{20} & 5.0\times 10^{-2} \\
H1821+643 & 94\fdg 00 & +27\fdg 42 & 8.1\times 10^{17} & 3.37\times 10^{-6} & 5.1\times 10^{20} & 3.2\times 10^{-3} \\
HE 1143$-$1810 & 281\fdg 85 & +41\fdg 71 & 3.5\times 10^{16} & 2.63\times 10^{-6} & 4.2\times 10^{20} & 1.7\times 10^{-4} \\
MRC 2251$-$178 & 46\fdg 20 & -61\fdg 33 & 3.5\times 10^{14} & 1.23\times 10^{-6} & 2.3\times 10^{20} & 3.0\times 10^{-6} \\
Mrk 9 & 158\fdg 36 & +28\fdg 75 & 2.3\times 10^{19} & 3.73\times 10^{-6} & 5.5\times 10^{20} & 8.4\times 10^{-2} \\
Mrk 335 & 108\fdg 76 & -41\fdg 42 & 6.8\times 10^{18} & 2.62\times 10^{-6} & 4.2\times 10^{20} & 3.2\times 10^{-2} \\
Mrk 509 & 35\fdg 97 & -29\fdg 86 & 7.4\times 10^{17} & 2.33\times 10^{-6} & 3.8\times 10^{20} & 3.9\times 10^{-3} \\
Mrk 1383 & 349\fdg 22 & +55\fdg 12 & 2.2\times 10^{14} & 1.55\times 10^{-6} & 2.8\times 10^{20} & 1.6\times 10^{-6} \\
Mrk 1513 & 63\fdg 67 & -29\fdg 07 & 2.6\times 10^{16} & 2.57\times 10^{-6} & 4.1\times 10^{20} & 1.3\times 10^{-4} \\
MS 0700.7+6338 & 152\fdg 47 & +25\fdg 63 & 5.6\times 10^{18} & 3.21\times 10^{-6} & 4.9\times 10^{20} & 2.3\times 10^{-2} \\
NGC 7469 & 83\fdg 10 & -45\fdg 47 & 4.7\times 10^{19} & 5.44\times 10^{-6} & 7.4\times 10^{20} & 1.3\times 10^{-1} \\
PG 0804+761 & 138\fdg 28 & +31\fdg 03 & 4.6\times 10^{18} & 2.58\times 10^{-6} & 4.1\times 10^{20} & 2.2\times 10^{-2} \\
PG 0844+349 & 188\fdg 56 & +37\fdg 97 & 1.7\times 10^{18} & 2.58\times 10^{-6} & 4.1\times 10^{20} & 8.2\times 10^{-3} \\
PG 1211+143 & 267\fdg 55 & +74\fdg 32 & 2.4\times 10^{18} & 1.82\times 10^{-6} & 3.2\times 10^{20} & 1.5\times 10^{-2} \\
PG 1302$-$102 & 308\fdg 59 & +52\fdg 16 & 4.2\times 10^{15} & 2.33\times 10^{-6} & 3.8\times 10^{20} & 2.2\times 10^{-5} \\
PKS 0558$-$504 & 257\fdg 96 & -28\fdg 57 & 2.8\times 10^{15} & 3.00\times 10^{-6} & 4.7\times 10^{20} & 1.2\times 10^{-5} \\
PKS 2155$-$304 & 17\fdg 73 & -52\fdg 25 & 1.5\times 10^{14} & 7.92\times 10^{-7} & 1.7\times 10^{20} & 1.8\times 10^{-6} \\
VII Zw 118 & 151\fdg 36 & +25\fdg 99 & 6.9\times 10^{18} & 3.24\times 10^{-6} & 4.9\times 10^{20} & 2.8\times 10^{-2} \\
\enddata
\tablecomments{
Columns (1): name of target, (2) and (3): position in the Galactic coordinates, (4): H$_{2}$ column density derived with the UV measurements \citep{2006ApJ...636..891G}, (5): dust optical depth at 353\,GHz \citep{2014A+A...571A..11P}, (6): total column density obtained from $\tau_{353}$ by taking into account a nonlinear relationship \citep{2017ApJ...838..132O}, (7): H$_{2}$ fraction given by $f_\mathrm{H2}=2N_\mathrm{H2}/(2N_\mathrm{H2}+N_{\text{H$\;$\sc{i}}})$.}
\end{deluxetable*}

\subsection{Synthetic observations of \ion{H}{1} Line Profiles}\label{ssec:syntheticobservations}
\subsubsection{Density distribution in a line of sight}
\begin{figure*}
\includegraphics{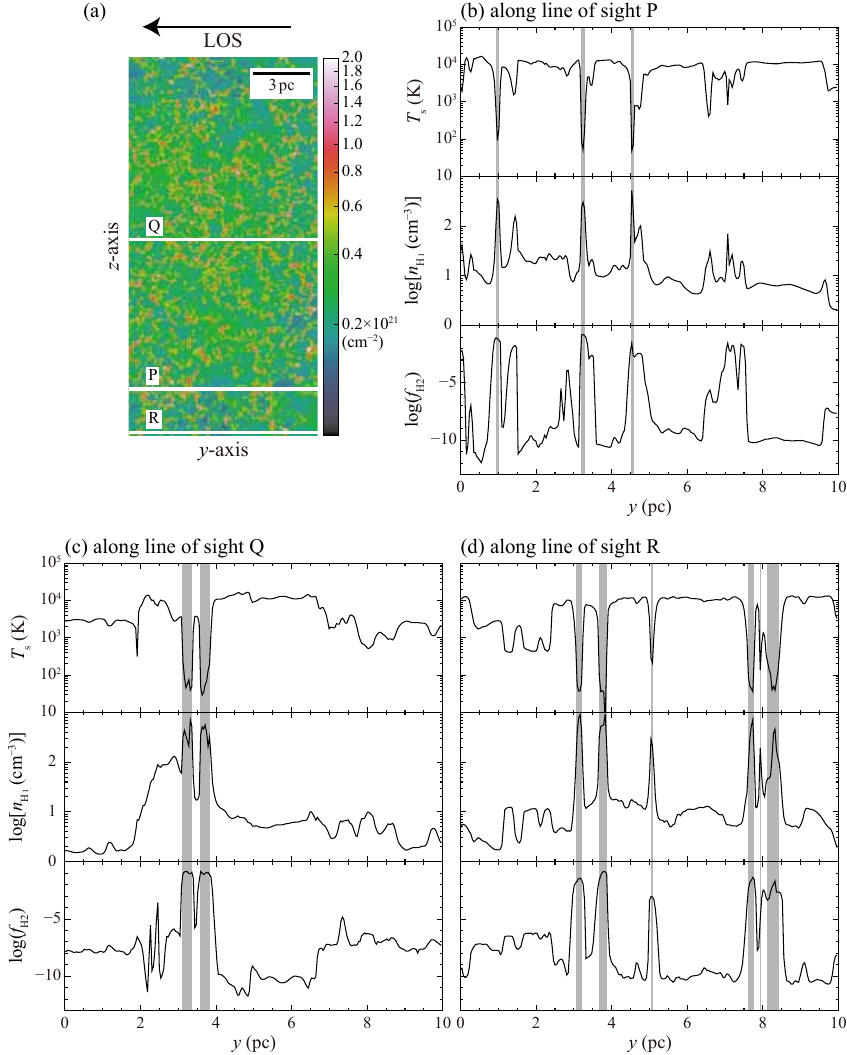}
\caption{
(a) Side view of the \ion{H}{1} column density in the $y$-$z$ plane integrated in the $x$-direction in the model.
Three lines of sight selected are shown by the lines P, Q and R (see the text and Table \ref{tab:WNMfraction}).
The image is $10\,\text{pc}\times 20\,\text{pc}$ in size.
The $y$ axis in the numerical domain corresponds to the horizontal axis and the $z$ axis to the vertical axis.
(b) $T_\mathrm{s}$, $n_{\text{H$\;$\sc{i}}}$ and H$_2$ fraction $f_\mathrm{H2}=2n_\mathrm{H2}/(2n_\mathrm{H2}+n_{\text{H$\;$\sc{i}}})$ profiles (from top to bottom) along the line of sight P.
The horizontal axis is the distance from the far-side of the model ISM, $y$.
The CNM spikes are indicated by shaded regions.
(c) and (d) Same as (b) but along lines of sight Q and R, respectively.
\label{fig:sideview}}
\end{figure*}

\begin{figure}
\includegraphics{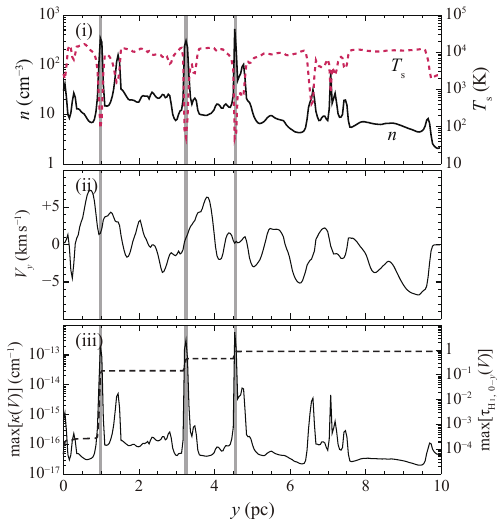}
\caption{
(i) Profiles of density (the solid black line) and spin temperature (the dashed red line) along the line of sight P.
The horizontal axis is the distance from the far-side of the model ISM, $y$.
Profiles of (ii) line-of-sight velocity, $V_{y}$, and (iii) peak opacity, $\max\left[\kappa(V)\right]$ given by Equation (\ref{eqn:kappa}), along the same line of sight as (i) are shown.
Peak optical depth $\max\left[\tau_{\text{H$\;$\sc{i}}, 0\text{--}y}(V)\right]=\max\left[\int_{0}^{y} \kappa(V) dy^\prime\right]$ is plotted against right-side vertical-axis of panel (iii) by the dashed line.
The CNM spikes are indicated by shaded regions.
\label{fig:linecalcparams}}
\end{figure}

Figure \ref{fig:sideview}(a) gives a side view of the density profile of the model in the $y$-$z$ plane integrated in the $x$-direction.
In order to show typical line profiles, we chose three lines of sight P, Q and R which have similar total 21\,cm line intensity $W_{\text{H$\;$\sc{i}}}$ with different column density $N_{\text{H$\;$\sc{i}}}$.
Panels (b)--(d) of Figure \ref{fig:sideview} show distributions of density and temperature for each pixel in the three lines of sight P, Q, and R.
The CNM appears as a few spikes with a sub-pc size whose typical density is $10^{2}$ to $10^{3}$\,cm$^{-3}$.
The WNM is distributed with density less than 100\,cm$^{-3}$ and show smooth distribution.

The distributions of various \ion{H}{1} physical parameters which are required in calculating line profiles are given in Figure \ref{fig:linecalcparams} for the line of sight P.
Figure \ref{fig:linecalcparams}(i) shows $n$ and $T_\mathrm{s}$, Figure \ref{fig:linecalcparams}(ii) line of sight velocity $V_\mathrm{y}$, and Figure \ref{fig:linecalcparams}(iii) the maximum opacity and the optical depth.
The opacity is integrated in the line of sight to yield optical depth.
The accumulated optical depth reaches $\sim 1$ and is determined by a few CNM spikes in each line of sight as seen in Figure \ref{fig:linecalcparams}(iii).
The WNM has little contribution to the optical depth, which is a natural consequence of the $T_\mathrm{s}^{-1}$-dependence of opacity (see Equation (\ref{eqn:kappa})).

\subsubsection{Calculations of \ion{H}{1} Line Profiles}
The simulated ISM is used to calculate synthetic \ion{H}{1} profiles by using the line radiation transfer equation given as
\begin{eqnarray}
I_{j+1}(V) & = & I_{j}(V)\exp\left[-\kappa_{j}(V)\Delta y\right]+\frac{\epsilon_{j}(V)}{\kappa_{j}(V)}\left\{1-\exp\left [-\kappa_j(V)\Delta y\right ]\right \},  \label{eqn:transfer}\\
I_0 & = & B(T_\mathrm{bg})\sim \frac{2\nu_0^{2}}{c^{2}}k_\mathrm{B}T_\mathrm{bg}, \label{eqn:transfer0}
\end{eqnarray}
where $I(V)$ the line intensity, $\kappa (V)$ opacity, $\epsilon (V)$ emissivity, $B(T)$ the Planck function at 21\,cm, $\nu_0=1.420405751$\,GHz, $T_\mathrm{bg}=2.7$\,K the brightness temperature of the background radiation field, $c$ the light velocity and $k_\mathrm{B}$ Boltzmann constant.
The subscript $j$ stands for the $j$-th cell along a line of sight.
The emissivity $\epsilon$ and opacity $\kappa$ of the 21\,cm transition at a radial velocity $V$ are calculated as follows;
\begin{eqnarray}
\epsilon_j(V) & = & \frac{hc}{4\pi}n_{\mathrm{up}, j}A\phi_j(V), \label{eqn:epsilon} \\
\kappa_j(V) & = & \frac{3c^3h}{8\pi\nu_0^2 k_\mathrm{B}T_{\mathrm{s}, j}}n_{\mathrm{low}, j}A\phi_j(V), \label{eqn:kappa}
\end{eqnarray}
where $h$ and $A=2.8688754\times 10^{-15}$\,s$^{-1}$ are the Planck constant and the Einstein $A$ coefficient, respectively.
The \ion{H}{1} optical depth at a radial velocity $V$, $\tau_{\text{H$\;$\sc{i}}}(V)$, obtained by integration in a line of sight is calculated as follows;
\begin{equation}\label{eqn:real_tau}
\tau_{\text{H$\;$\sc{i}}}(V)=\sum_{j} \left[\kappa_{j}(V)\Delta y\right].
\end{equation}

The number density of H atom in the lower state is given by
\begin{equation}\label{eqn:n_low}
n_{\mathrm{low}, j}=\frac{n_{\text{H$\;$\sc{i}}, j}}{3\exp\left(\frac{\displaystyle -h\nu_0}{\displaystyle k_\mathrm{B}T_{\mathrm{s}, j}}\right)+1}
\end{equation}
and that in the upper state is 
\begin{equation}\label{eqn:n_up}
n_{\mathrm{up}, j}=n_{\text{H$\;$\sc{i}}, j} - n_{\mathrm{low}, j}
\end{equation}
for total \ion{H}{1} density $n_{\text{H$\;$\sc{i}}}$.
The line shape function
\begin{equation}\label{eqn:distributionfunc}
\phi_j(V)=\sqrt{\frac{m_\mathrm{p}+m_\mathrm{e}}{2k_\mathrm{B}T_{\mathrm{s}, j}\pi}}\exp\left[\frac{-(m_\mathrm{p}+m_\mathrm{e})(V-V_{y, j})^{2}}{2k_\mathrm{B}T_{\mathrm{s}, j}}\right]
\end{equation}
satisfies $\int \phi(V) dV=1$, where $m_\mathrm{p}=1.67262178\times 10^{-24}$\,g is the mass of a proton and $m_\mathrm{e}=9.10938291\times 10^{-28}$\,g is that of an electron.
The \ion{H}{1} spin temperature $T_\mathrm{s}$ is derived by applying a method of \citet{2014ApJ...786...64K}, which gives $T_\mathrm{s}\sim T_\mathrm{k}$ in a $T_\mathrm{k}$ range from 20 to $3\times 10^{3}$\,K.
For $T_\mathrm{k} < 20$\,K, we simply adopt $T_\mathrm{s}=T_\mathrm{k}$.
The $\sim$75\% of the data pixels have $T_\mathrm{s}/T_\mathrm{k}=0.9$--1.0 and the others $T_\mathrm{s}/T_\mathrm{k}=0.8$--0.9.

\subsubsection{Emission profiles}
An \ion{H}{1} line profile is calculated by integrating the line transfer Equations (\ref{eqn:transfer}) and (\ref{eqn:transfer0}) from the far side to the near side of the data cube along the $y$-axis seen by the observer over a distance of 10\,pc, a half of the full span of the data cube, and the observed brightness temperature is given by
\begin{equation}
T_\mathrm{b}(V)=I(V)\frac{c^{2}}{2\nu_0^{2}k_\mathrm{B}}-T_\mathrm{bg},
\end{equation}
which is approximated for convenience as
\begin{equation}
T_\mathrm{b}(V)=\left[T_\mathrm{s}(V)-T_\mathrm{bg}\right]\left\{1-\exp\left[-\tau_{\text{H$\;$\sc{i}}}(V)\right]\right\},
\end{equation}
where $T_\mathrm{s}$ is a harmonic mean spin temperature in a line of sight,
\begin{equation}
\frac{\sum_{j}\left[n_{\text{H$\;$\sc{i}},j}\phi_{j}(V)\right]}{T_\mathrm{s}(V)}=\sum_{j}\left[\frac{n_{\text{H$\;$\sc{i}}, j}\phi_{j}(V)}{T_{\mathrm{s}, j}}\right]
\end{equation}
and $T_\mathrm{bg}$ is subtracted as in real observations.

\begin{figure*}
\includegraphics{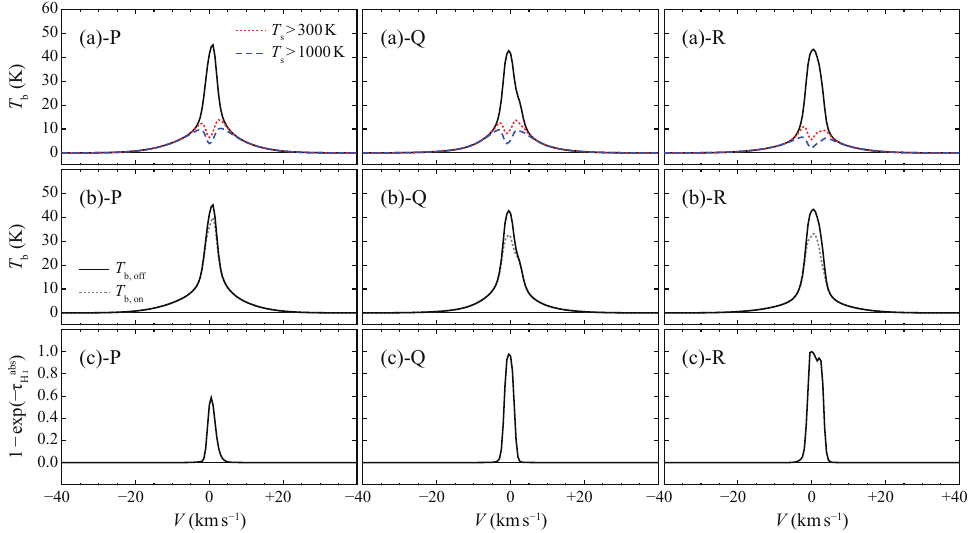}
\caption{
(a) Synthetic-observed \ion{H}{1} emission spectra toward the different three directions.
The solid lines are the emission of the whole \ion{H}{1} gas, and the short-dashed- and long-dashed-lines give the emission from the warm gas with $T_\mathrm{s}$ higher than 300\,K and 1000\,K, respectively.
The profiles only from the warm gas are calculated by setting the emissivity $\epsilon$ of the cold gas equal to 0, while the opacity $\kappa$ of the whole gas is held fixed.
(b) Synthetic-observed \ion{H}{1} emission spectra. The solid lines show $T_\mathrm{b, off}$ (Equation (\ref{eqn:tboff})) and the gray dashed lines show $T_\mathrm{b, on}$ (Equation (\ref{eqn:tbon})).
Here we assume a model background continuum source with a flux of 1\,Jy.
(c) Absorption spectra given by Equation (\ref{eqn:exptauabs}).
\label{fig:samplespectra}}
\end{figure*}

Figure \ref{fig:samplespectra}(a) shows three emission line profiles; the black solid line is the emission of the whole \ion{H}{1} gas, and the short-dashed- and long-dashed-lines give the emission from the warm gas with $T_\mathrm{s}$ higher than 300\,K and 1000\,K, respectively.
$T_\mathrm{s} > 1000$\,K is shown for reference.
The profiles only from the warm gas are calculated by setting the emissivity $\epsilon$ of the cold gas equal to 0, while the opacity $\kappa$ of the whole gas is fixed.
It is seen that a $T_\mathrm{s}$ value is not so critical in discriminating the CNM and WNM.

Figure \ref{fig:samplespectra}(a)-P shows an \ion{H}{1} emission profile calculated for the line of sight P in Figure 5 for 4{\arcmin} resolution, the same with the Arecibo telescope.
The intensity integrated over velocity gives the total intensity of a 21\,cm line emission profile $W_{\text{H$\;$\sc{i}}}$.
The CNM has a narrow profile whose linewidth is $\sim 5$\,km\,s$^{-1}$, while the WNM shows a broad wing-like profile of $\sim40$\,km\,s$^{-1}$ velocity span.
The profiles are consistent with the observed one in the solar neighborhood \citep[see the \ion{H}{1} profiles at high $b$, e.g.,][]{2005A&A...440..775K,2014ApJ...796...59F,2015ApJ...798....6F}.
The model allows us to separate the contributions of the CNM and WNM, which is impossible in real observations.
Figures \ref{fig:samplespectra}(a)-Q and \ref{fig:samplespectra}(a)-R show similar profiles in Q and R.
Table \ref{tab:WNMfraction} lists the derived physical parameters of the \ion{H}{1} gas. 

\subsubsection{Emission-absorption profiles and \ion{H}{1} column density}
The emission-absorption method uses absorption on a radio continuum source and averaged emission profiles off the continuum source by assuming that the \ion{H}{1} emission profiles are  not significantly different between the two positions.
An \ion{H}{1} emission profile is a function of two unknowns, $T_\mathrm{s}(V)$ and $\tau_{\text{H$\;$\sc{i}}}(V)$, for a single observed quantity $T_\mathrm{b}(V)$, and cannot be solved for $T_\mathrm{s}(V)$ and $\tau_{\text{H$\;$\sc{i}}}(V)$.
By observing a radio continuum source as a background source, a second equation which relates $T_\mathrm{s}(V)$ and $\tau_{\text{H$\;$\sc{i}}}(V)$ is obtained, and the two equations are coupled to derive $T_\mathrm{s}(V)$ and $\tau_{\text{H$\;$\sc{i}}}(V)$ \citep[e.g.,][]{2011piim.book.....D}.

The off-source brightness temperature $T_\mathrm{b, off}$ and on-source brightness temperature $T_\mathrm{b, on}$ are given as follows;
\begin{equation}
T_\mathrm{b, off}(V)=\left[T_\mathrm{s}(V)-2.7\,\text{K}\right]\left\{1-\exp\left[-\tau_{\text{H$\;$\sc{i}}}(V)\right]\right\},\label{eqn:tboff}
\end{equation}
and
\begin{equation}
T_\mathrm{b, on}(V)=\left[T_\mathrm{s}(V)-T_\mathrm{cont}\right]\left\{1-\exp\left[-\tau_{\text{H$\;$\sc{i}}}(V)\right]\right\}.\label{eqn:tbon}
\end{equation}
Here $T_\mathrm{cont}$ is the temperature of an assumed background compact continuum source.
The absorption spectra obtained from the emission-absorption measurements are given as,
\begin{equation}
1-\exp\left[-\tau_{\text{H$\;$\sc{i}}}^\mathrm{abs}(V)\right]=\frac{T_\mathrm{b, off}(V)-T_\mathrm{b, on}(V)}{T_\mathrm{cont}-2.7\,\text{K}}. \label{eqn:exptauabs}
\end{equation}

Figure \ref{fig:samplespectra}(b)-P shows $T_\mathrm{b, off}(V)$ in Equation (\ref{eqn:tboff}) (solid line) and $T_\mathrm{b, on}(V)$ in Equation (\ref{eqn:tbon}) (dashed line) toward a radio continuum compact source at P for 4{\arcmin} resolution and Figure \ref{fig:samplespectra}(c)-P shows $1-\exp\left[-\tau_{\text{H$\;$\sc{i}}}^\mathrm{abs}(V)\right]$ in Equation (\ref{eqn:exptauabs}).
The angular size of the radio continuum source is assumed to be equivalent to the pixel size 0{\farcm}9, which is nearly consistent with the typical size of the radio continuum sources 20{\arcsec}--30{\arcsec}.
In the real emission-absorption measurements the off-source spectrum is taken with a larger beam than the size of the radio continuum compact source.
Figures \ref{fig:samplespectra}(a)--(c)-Q and \ref{fig:samplespectra}(a)--(c)-R are the same profiles for the directions Q and R. 

In real observations an average of profiles near the on source position is used as $T_\mathrm{b, off }$ \citep[e.g.,][]{2003ApJS..145..329H}.
In the synthetic observations we use the on-source emission profile by assuming a model background continuum source with a flux density of 1\,Jy.
As readily confirmed $\tau_{\text{H$\;$\sc{i}}}^\mathrm{abs}(V)$ is equal to $\tau_{\text{H$\;$\sc{i}}}^\mathrm{model}(V)=\int \kappa(V)dy$ integrated in the line of sight. 

\section{OBSERVED PROPERTIES OF THE \ion{H}{1} GAS}\label{sec:observedprops}
\begin{figure*}
\includegraphics{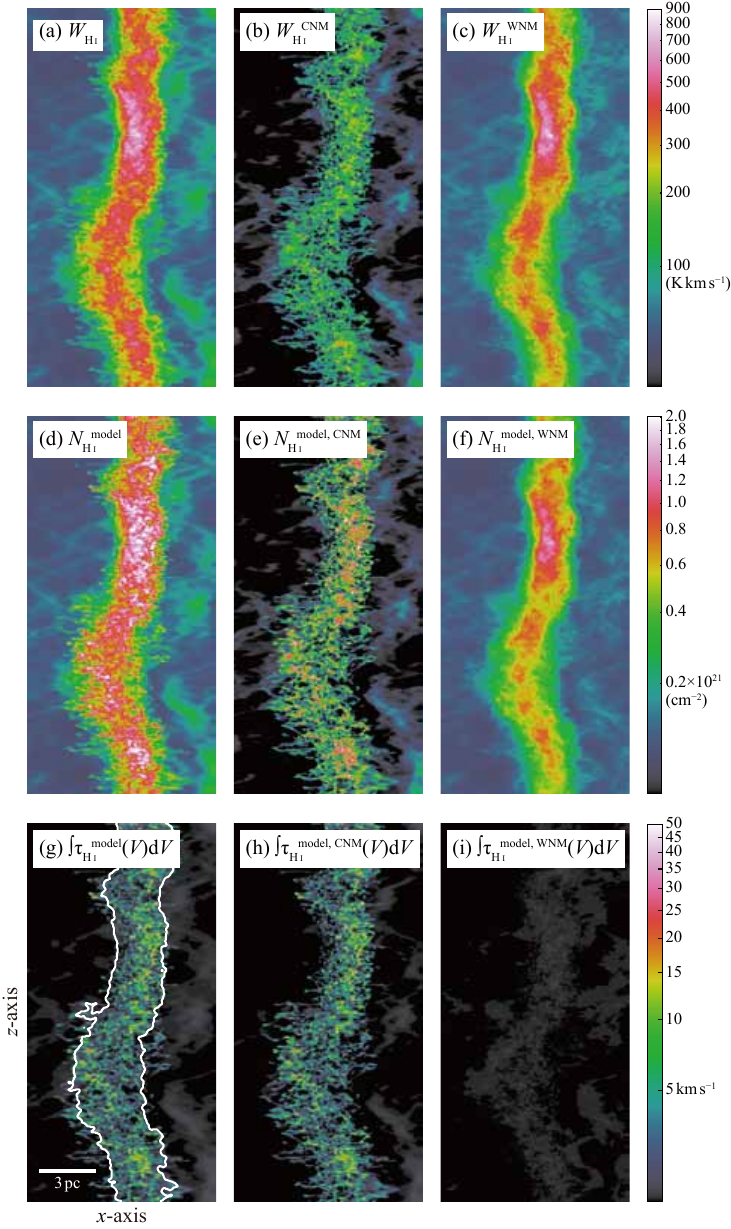}
\caption{
Spatial distribution of the 0.5-Myr model, (a) velocity-integrated intensity of the synthetic-observed \ion{H}{1} spectra ($W_{\text{H$\;$\sc{i}}}$), (b) $W_{\text{H$\;$\sc{i}}}$ produced from the CNM with $T_\mathrm{k}<300$\,K, (c) $W_{\text{H$\;$\sc{i}}}$ produced from the WNM with $T_\mathrm{k}>300$\,K, (d) model \ion{H}{1} column density ($N_{\text{H$\;$\sc{i}}}^\mathrm{model}$), (e) CNM column density, (f) WNM column density, (g) velocity-integrated model optical depth ($\int \tau_{\text{H$\;$\sc{i}}}^\mathrm{model}(V)dV$), (h) those produced from the CNM and (i) produced from the WNM in the 0.5-Myr model.
The images are $10\,\text{pc}\times 20\,\text{pc}$ in size and have a resolution of 0.04\,pc per pixel.
The $x$ axis in the numerical domain corresponds to the horizontal axis of each panel and the $z$ axis to the vertical axis.
The contor in panel (g) outlines the ROI (identical to that shown in Figure \ref{fig:v-rendering}).
\label{fig:maps}}
\end{figure*}

\begin{figure}
\includegraphics{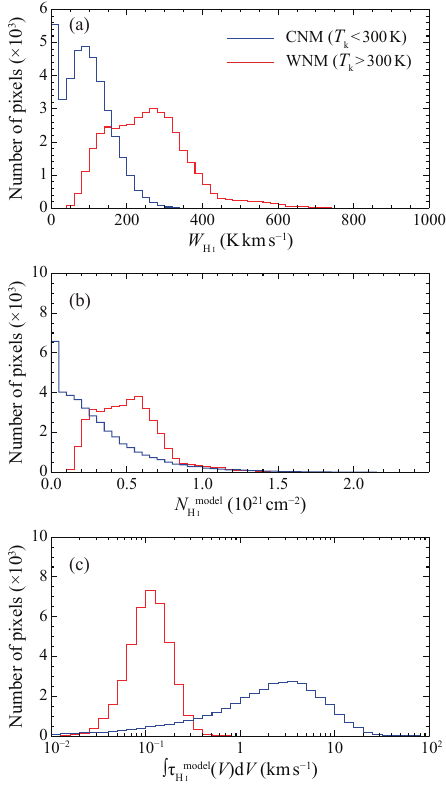}
\caption{
Histograms of (a) $W_{\text{H$\;$\sc{i}}}$, (b) $N_{\text{H$\;$\sc{i}}}^\mathrm{model}$ and (c) $\int \tau_{\text{H$\;$\sc{i}}}^\mathrm{model}(V)dV$ in the 0.5-Myr model.
The blue lines represent the contribution of the CNM and red lines represent that of the WNM.
\label{fig:histo_W_N_tau}}
\end{figure}

Synthetic observations provide total \ion{H}{1} 21\,cm line intensity $W_{\text{H$\;$\sc{i}}}$.
Other parameters obtained include the \ion{H}{1} column density $N_{\text{H$\;$\sc{i}}}$, and the \ion{H}{1} optical depth $\tau_{\text{H$\;$\sc{i}}}$.
We calculated these \ion{H}{1} parameters for the CNM and the WNM separately.
The projected distributions of $W_{\text{H$\;$\sc{i}}}$, $N_{\text{H$\;$\sc{i}}}^\mathrm{model}$ and velocity-integrated $\tau_{\text{H$\;$\sc{i}}}^\mathrm{model}$, $\int \tau_{\text{H$\;$\sc{i}}}^\mathrm{model}(V)dV$ are shown in the nine panels of Figures \ref{fig:maps}(a)--(i).
We note significant difference between the CNM and the WNM.
The CNM is highly filamentary and the WNM shows smooth distribution in the three parameters.

Figure \ref{fig:histo_W_N_tau} shows three histograms of $W_{\text{H$\;$\sc{i}}}$, $N_{\text{H$\;$\sc{i}}}^\mathrm{model}$ and $\int \tau_{\text{H$\;$\sc{i}}}^\mathrm{model}(V)dV$ for the CNM and the WNM, respectively.
A clear trend is that $N_{\text{H$\;$\sc{i}}}^\mathrm{model}$ is nearly comparable between the CNM and the WNM (Figure \ref{fig:histo_W_N_tau}(b)).
$\int \tau_{\text{H$\;$\sc{i}}}^\mathrm{model}(V)dV$ is dominated exclusively by the optically thick CNM (Figure \ref{fig:histo_W_N_tau}(c)).
These properties are shown numerically in Table 5.
\added{
Table \ref{tab:WNMfraction} shows details of the \ion{H}{1} parameters including $W_{\text{H$\;$\sc{i}}}$, $N_{\text{H$\;$\sc{i}}}$, $T_\mathrm{s}$, and $\int \tau_{\text{H$\;$\sc{i}}}^\mathrm{model}dV$ for representative three line profiles, which have similar $W_{\text{H$\;$\sc{i}}}$.
$N_{\text{H$\;$\sc{i}}}^\mathrm{model}$ is the integrated column density of the model and $N_{\text{H$\;$\sc{i}}}^\mathrm{thin}$ is calculated from $W_{\text{H$\;$\sc{i}}}$ by Equation (\ref{eqn:opticallythin_eqn}).
$W_{\text{H$\;$\sc{i}}}^\mathrm{WNM}$ is larger than the $W_{\text{H$\;$\sc{i}}}^\mathrm{CNM}$ at lower $N_{\text{H$\;$\sc{i}}}$, while $W_{\text{H$\;$\sc{i}}}^\mathrm{CNM}$ becomes comparable to $W_{\text{H$\;$\sc{i}}}^\mathrm{WNM}$ at higher $N_{\text{H$\;$\sc{i}}}$.
$W_{\text{H$\;$\sc{i}}}^\mathrm{WNM}$ becomes large because $T_\mathrm{s}$ is high, in spite of the small $\int \tau_{\text{H$\;$\sc{i}}}(V)dV$.
The CNM is usually optically thick with $\int \tau_{\text{H$\;$\sc{i}}}(V)dV$ larger than $\sim 2$, while the WNM is always optically thin.
This large optical depth of the CNM make the ratio $N_{\text{H$\;$\sc{i}}}^\mathrm{model}/N_{\text{H$\;$\sc{i}}}^\mathrm{thin}$ larger than 1.0 at higher $\int \tau_{\text{H$\;$\sc{i}}}(V)dV$.
The WNM shows similar $N_{\text{H$\;$\sc{i}}}^\mathrm{model}$ and $N_{\text{H$\;$\sc{i}}}^\mathrm{thin}$, while $N_{\text{H$\;$\sc{i}}}^\mathrm{thin}$ is by a factor of 1.1--1.2 smaller than $N_{\text{H$\;$\sc{i}}}^\mathrm{model}$.
This is due the absorption by the CNM.
$\langle T_\mathrm{s} \rangle$ represents a density-weighted harmonic mean of $T_\mathrm{s}$ in the line of sight expressed as follows;
}
\begin{equation}
\frac{\sum_{j}(n_{\text{H$\;$\sc{i}}, j})}{\langle T_\mathrm{s}\rangle}=\sum_{j}\left(\frac{n_{\text{H$\;$\sc{i}}, j}}{T_{\mathrm{s}, j}}\right).\label{eqn:harmonicmean_ts}
\end{equation}
\added{
Equation (\ref{eqn:harmonicmean_ts}) reflects that $\int \tau_{\text{H$\;$\sc{i}}}(V)dV$ is the sum of the contribution of different $T_\mathrm{s}$ components in the line of sight.
$\langle T_\mathrm{s}\rangle$ is calculated for the CNM and WNM, respectively, as well as for the whole profile.
We note that the velocity averaged $T_\mathrm{s}$ in \citet{2014ApJ...796...59F,2015ApJ...798....6F} corresponds to $\langle T_\mathrm{s}\rangle$ for the whole in the present notation.
}
It is notable that the WNM shows practically no contribution to $\tau_{\text{H$\;$\sc{i}}}^\mathrm{model}$ (Figure \ref{fig:histo_W_N_tau}(c)).

\begin{figure*}
\includegraphics{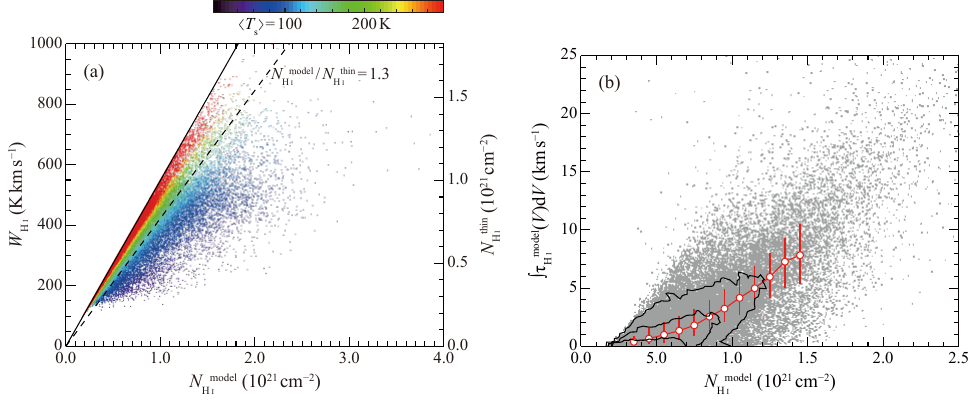}
\caption{
(a) Correlation plot of $W_{\text{H$\;$\sc{i}}}$ versus $N_{\text{H$\;$\sc{i}}}^\mathrm{model}$.
Color represents $\langle T_\mathrm{s}\rangle $ of each point.
The right-sight vertical axis shows the column density given assuming optically thin approximation, $N_{\text{H$\;$\sc{i}}}^\mathrm{thin}=1.823\times 10^{18}$\,(cm$^{-2}$\,K$^{-1}$\,km$^{-1}$\,s)\,$W_{\text{H$\;$\sc{i}}}$.
The solid and dashed lines indicate optically thin limit ($N_{\text{H$\;$\sc{i}}}^\mathrm{model}/N_{\text{H$\;$\sc{i}}}^\mathrm{thin}=1.0$) and $N_{\text{H$\;$\sc{i}}}^\mathrm{model}/N_{\text{H$\;$\sc{i}}}^\mathrm{thin}=1.3$, respectively.
(b) Scatter plot between $N_{\text{H$\;$\sc{i}}}^\mathrm{model}$ and $\int \tau_{\text{H$\;$\sc{i}}}^\mathrm{model}(V)dV$.
The circles and vertical bars show the median and interquartile range of $\int \tau_{\text{H$\;$\sc{i}}}^\mathrm{model}(V)dV$ in each $1 \times 10^{20}$\,cm$^{-2}$ bin.
\label{fig:NHI-WHI}}
\end{figure*}

Figure \ref{fig:NHI-WHI}(a) show a scatter plot between $W_{\text{H$\;$\sc{i}}}$ and $N_{\text{H$\;$\sc{i}}}^\mathrm{model}$, where $\langle T_\mathrm{s}\rangle$ is indicated in a color code and $N_{\text{H$\;$\sc{i}}}^\mathrm{model}/N_{\text{H$\;$\sc{i}}}^\mathrm{thin}=1.3$ by the dashed line.
At $\langle T_\mathrm{s}\rangle$ higher than 200\,K, the optically thin approximation produces a linear relationship between $W_{\text{H$\;$\sc{i}}}$ and $N_{\text{H$\;$\sc{i}}}^\mathrm{model}$ as expressed by Equation (\ref{eqn:opticallythin_eqn}), whereas at $\langle T_\mathrm{s}\rangle$ lower than 100\,K the \ion{H}{1} optical depth becomes larger and $W_{\text{H$\;$\sc{i}}}$ becomes weaker than the thin limit due to saturation.
\added{This is consistent with the optically thick \ion{H}{1} derived from the \textit{Planck}/\textit{IRAS}-based analysis by \citet{2015ApJ...798....6F}}.
Figure \ref{fig:NHI-WHI}(b) shows $\int \tau_{\text{H$\;$\sc{i}}}^\mathrm{model}(V)dV$ as a function of $N_{\text{H$\;$\sc{i}}}^\mathrm{model}$.
The median solid curve indicates that $\int \tau_{\text{H$\;$\sc{i}}}^\mathrm{model}(V)dV$ increases rapidly with $N_{\text{H$\;$\sc{i}}}^\mathrm{model}$ like $(N_{\text{H$\;$\sc{i}}}^\mathrm{model})^2$.
This is explained by the relationship $\tau_{\text{H$\;$\sc{i}}}=\text{constant}\times N_{\text{H$\;$\sc{i}}}/(T_\mathrm{s}\Delta V)$ (see Equation \ref{eqn:exptauabs}) where $T_\mathrm{s}$ is proportional to $n^{-1}$ (Figure \ref{fig:histo_n_T}(c)) if $\Delta V$ remains fixed.

\begin{deluxetable*}{lRRRRRR}
\tablecolumns{7}
\tablewidth{0pc}
\tablecaption{Physical parameters of the three samples of \ion{H}{1} profiles in Figure \ref{fig:samplespectra}\label{tab:WNMfraction}}
\tablehead{
 & \colhead{$W_{\text{H$\;$\sc{i}}}$} & \colhead{$N_{\text{H$\;$\sc{i}}}^\mathrm{model}$} & \colhead{$N_{\text{H$\;$\sc{i}}}^\mathrm{thin}$} & \colhead{$N_{\text{H$\;$\sc{i}}}^\mathrm{model}/N_{\text{H$\;$\sc{i}}}^\mathrm{thin}$} & \colhead{$\langle T_\mathrm{s}\rangle$} & \colhead{$\int \tau_{\text{H$\;$\sc{i}}}^\mathrm{model}(V)dV$} \\
 & \colhead{(K\,km\,s$^{-1}$)} & \colhead{($10^{20}$\,cm$^{-2}$)} & \colhead{($10^{20}$\,cm$^{-2}$)} & & \colhead{(K)} &\colhead{(km\,s$^{-1}$)}   
}
\colnumbers
\startdata
\multicolumn{7}{l}{Sample P (Figure \ref{fig:samplespectra}(a)--(c)-P)} \\
\hline
whole &  354 & 7.4 & 6.5 & 1.1 & 200 & 2.0 \\
WNM & 223 & 4.8 & 4.1 & 1.2 & 2505 & 0.11  \\
CNM & 131 & 2.6 & 2.4 & 1.1 & 73 & 1.9 \\
\hline
\multicolumn{7}{l}{Sample Q (Figure \ref{fig:samplespectra}(a)--(c)-Q)} \\
\hline
whole & 353 & 10.7 & 6.4 & 1.7 & 82 & 7.2 \\
WNM & 198 & 3.9 & 3.6 & 1.1 & 3523 & 0.06 \\
CNM & 155 & 6.8 & 2.8 & 2.4 & 53 & 7.1 \\
\hline
\multicolumn{7}{l}{Sample R (Figure \ref{fig:samplespectra}(a)--(c)-R)} \\
\hline
whole &  356 & 14.5 & 6.4 & 2.3 & 33 & 25.0 \\
WNM & 158 & 3.4 & 2.9 & 1.2 & 1136 & 0.17 \\
CNM & 198 & 11.1 & 3.6 & 3.1 & 25 & 24.8 \\
\enddata
\tablecomments{
Columns (2): velocity-integrated intensity, (3): column density of the model, (4): column density obtained under the optically-thin assumption (Equation (\ref{eqn:opticallythin_eqn})), (5): ratio of (3) to (4), (6): density-weighted harmonic mean of $T_\mathrm{s}$ along the lines-of-sight (Eq.\ \ref{eqn:harmonicmean_ts}), (7): velocity-integrated model optical-depth.}
\end{deluxetable*}

\section{DISCUSSION; \ion{H}{1} FILAMENTARY DISTRIBUTION AND ITS IMPACT ON THE EMISSION-ABSORPTION MEASUREMENTS}\label{sec:discussion}

\subsection{The CNM Filaments Observed in the Emission-Absorption Measurements}\label{ssec:cnmfilaments}
\begin{figure*}
\includegraphics{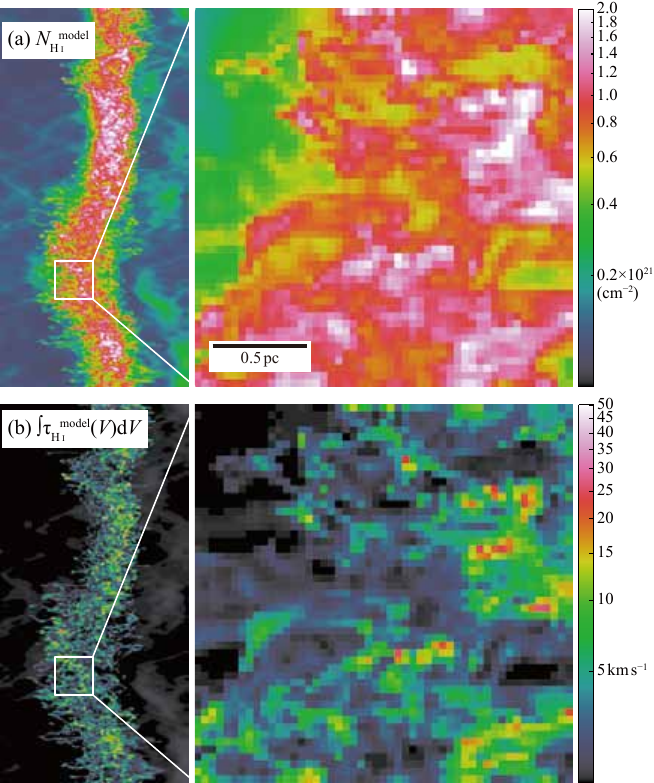}
\caption{
(a) The left panel shows spatial distribution of $N_{\text{H$\;$\sc{i}}}^\mathrm{model}$ (identical to Figure \ref{fig:maps}(d)), and the right panel shows a close-up view of the bounding box overlaid on the left panel.
(b) Same as (a) but for $\int \tau_{\text{H$\;$\sc{i}}}^\mathrm{model}(V)dV$.
\label{fig:closeup}}
\end{figure*}

Figures \ref{fig:closeup}(a) and \ref{fig:closeup}(b) show detailed enlarged distributions of $N_{\text{H$\;$\sc{i}}}^\mathrm{model}$ and $\int \tau_{\text{H$\;$\sc{i}}}^\mathrm{model}(V)dV$, respectively.
Because of the non-linear behavior of $\int \tau_{\text{H$\;$\sc{i}}}(V)dV$, which is dominated by the CNM (Figure \ref{fig:NHI-WHI}(b)), $\int \tau_{\text{H$\;$\sc{i}}}(V)dV$ enhances the filamentary CNM distribution.
The typical width of the filaments is less than 0.1\,pc with their length of roughly 0.5\,pc or more.
It is conceivable that the small scale structures of $\int \tau_{\text{H$\;$\sc{i}}}(V)dV$ significantly affect high resolution observations, and we explore the resolution effects in \ion{H}{1} observations below.

\begin{figure}
\includegraphics{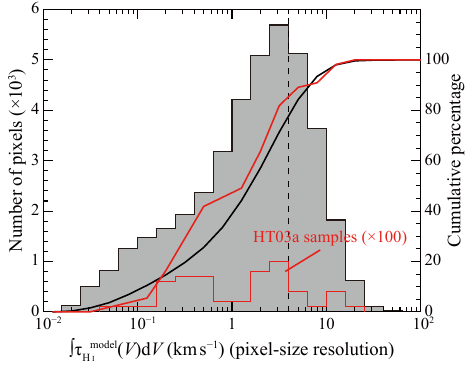}
\caption{
Histogram of $\int \tau_{\text{H$\;$\sc{i}}}^\mathrm{model}(V)dV$ at pixel-sized resolution.
The dashed line shows $\int \tau_{\text{H$\;$\sc{i}}}^\mathrm{model}(V)dV=4$\,km\,s$^{-1}$ and the black solid curve shows cumulative percentage plotted against right-side vertical-axis.
Histogram and cumulative percentage of $\int \tau_{\text{H$\;$\sc{i}}}^\mathrm{abs}(V)dV$ of \citet{2003ApJS..145..329H} samples with $|b|>15\arcdeg$ (Table \ref{tab:HTsamples}) are overlaid by the red lines (the former is multiplied by a factor of 100).
\label{fig:pix-4arcmin}}
\end{figure}

\subsubsection{Observed values of $\tau_{\text{H$\;$\sc{i}}}$}\label{sssec:obsvaluestauHI}
Figure \ref{fig:pix-4arcmin} shows a histogram of $\int \tau_{\text{H$\;$\sc{i}}}^\mathrm{model}(V)dV$ at pixel-sized resolution of the spatial distribution presented in Figure \ref{fig:maps}(g).
$\int \tau_{\text{H$\;$\sc{i}}}^\mathrm{model}(V)dV$ is distributed over a wide range from $2.5\times 10^{-2}$\,km\,s$^{-1}$ to $\sim 25$\,km\,s$^{-1}$ at a 5\% level of the histogram, and from $6\times 10^{-2}$\,km\,s$^{-1}$ to 16\,km\,s$^{-1}$ at a 20\% level.
We also note that the small $\int \tau_{\text{H$\;$\sc{i}}}^\mathrm{model}(V)dV$ tail in $\int \tau_{\text{H$\;$\sc{i}}}^\mathrm{model}(V)dV$ below 1\,km\,s$^{-1}$ is significant, reflecting the wide spread WNM with low $\tau_{\text{H$\;$\sc{i}}}$.
Conversely, high $\tau_{\text{H$\;$\sc{i}}}$ points at $\int \tau_{\text{H$\;$\sc{i}}}^\mathrm{model}(V)dV$ more than 4\,km\,s$^{-1}$ is dominated by the compact CNM.

Pixel-sized resolution measurements are carried out in the emission-absorption measurements toward radio continuum compact sources, where $\int \tau_{\text{H$\;$\sc{i}}}^\mathrm{abs}(V)dV$ is often used as an observable quantity which characterizes \ion{H}{1} gas property in the previous papers \citep[e.g.,][]{2014ApJ...793..132S}.
The effective resolution in the absorption measurements is given by the size of the radio continuum source and is typically $\lesssim 20{\arcsec}$--30{\arcsec} \citep{2003ApJS..145..329H,2014ApJ...793..132S}, nearly consistent with the present pixel size 0{\farcm}9.
The number density of radio sources in the published measurements is small.
In the Perseus region \citep{2014ApJ...793..132S,2015ApJ...809...56L}, the number of radio continuum sources is 27 for 500 square degrees, and the source density in \citet{2003ApJS..145..329H} is similar to that (see Table \ref{tab:HTsamples}).
This indicates source density 0.05\,deg$^{-2}$ or 0.005\,pc$^{-2}$ at 200\,pc in the sky, and corresponds to about 0.25 sources in the present \ion{H}{1} distribution having $\sim 50$\,pc$^{2}$.
The fraction of the sky measured in the emission-absorption measurements is therefore as small as $\sim 4\times 10^{-6}$ if a source diameter is assumed to be 30{\arcsec} or 0.03\,pc at 200\,pc. 

The present synthetic observations show that the emission-absorption measurements toward the present model \ion{H}{1} gas will obtain $\int \tau_{\text{H$\;$\sc{i}}}(V)dV$ whose probability distribution is given by Figure \ref{fig:pix-4arcmin}.
The measurements will find smaller $\int \tau_{\text{H$\;$\sc{i}}}(V)dV$ less than 4\,km\,s$^{-1}$ at a probability of 70\% and less than 1\,km\,s$^{-1}$ at a probability of 40\%.
Conversely, it is possible that $\int \tau_{\text{H$\;$\sc{i}}}(V)dV$ higher than 10\,km\,s$^{-1}$ is obtained at a probability of 5\% toward peaks of the CNM.
So, the general trend observed in the emission-absorption measurements is ``smaller $\int \tau_{\text{H$\;$\sc{i}}}(V)dV$'' of $10^{-1}$\,km\,s$^{-1}$--10\,km\,s$^{-1}$ with a large dispersion over two orders of magnitude at a 20\% level in Figure \ref{fig:pix-4arcmin}.
It is a question how well the measured $\tau_{\text{H$\;$\sc{i}}}$ in the emission-absorption measurements represents the \ion{H}{1} gas property given the extremely small source number density.
The usual assumption of uniform \ion{H}{1} gas in the emission-absorption measurements is far from reality in the present model. 
It is also to be noted that $T_\mathrm{s}(V)$ and $\tau_{\text{H$\;$\sc{i}}}(V)$ differs generally in between equations (\ref{eqn:tboff}) and (\ref{eqn:tbon}) contrary to the assumption of the emission-absorption measurements.

\begin{figure}
\includegraphics{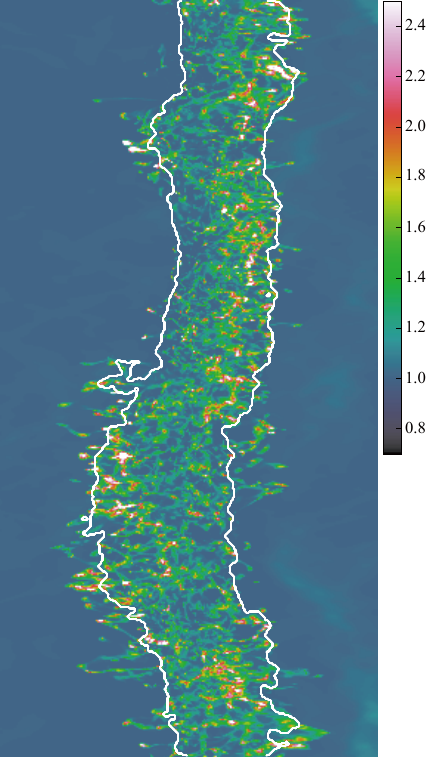}
\caption{
Spatial distribution of the $N_{\text{H$\;$\sc{i}}}^\mathrm{model}/N_{\text{H$\;$\sc{i}}}^\mathrm{thin}$ ratio at pixel-sized resolution.
The contour outlines the ROI (identical to that shown in Figures \ref{fig:v-rendering} and \ref{fig:maps}(g)).
\label{fig:model_thin_map}}
\end{figure}

\begin{figure*}
\includegraphics{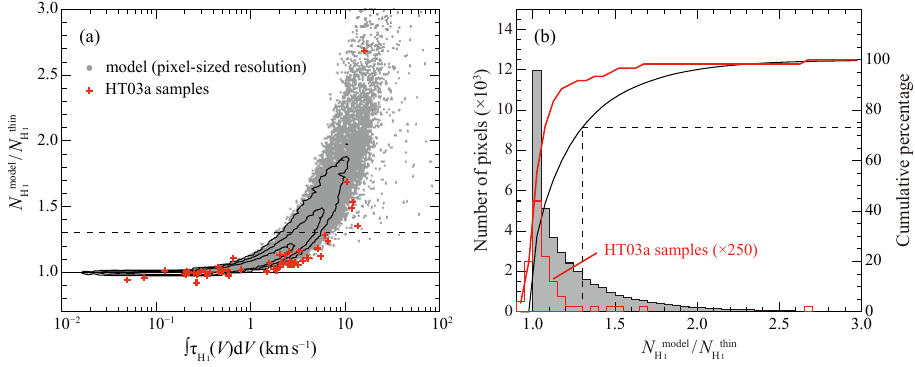}
\caption{
(a) Scatter plot between $\int \tau_{\text{H$\;$\sc{i}}}^\mathrm{model}(V)dV$ and $N_{\text{H$\;$\sc{i}}}^\mathrm{model}/N_{\text{H$\;$\sc{i}}}^\mathrm{thin}$ at pixel-sized resolution.
Contours include 50\%, 75\% and 90\% of data points.
The solid and dashed lines show $N_{\text{H$\;$\sc{i}}}^\mathrm{model}/N_{\text{H$\;$\sc{i}}}^\mathrm{thin}=1.0$ and 1.3, respectively.
The red crosses show scatter plot between $\int \tau_{\text{H$\;$\sc{i}}}^\mathrm{abs}(V)dV$ and $N_{\text{H$\;$\sc{i}}}^\mathrm{HT}/N_{\text{H$\;$\sc{i}}}^\mathrm{thin}$ for \citet{2003ApJS..145..329H} samples (table \ref{tab:HTsamples}). 
(b) Histogram of $N_{\text{H$\;$\sc{i}}}^\mathrm{model}/N_{\text{H$\;$\sc{i}}}^\mathrm{thin}$ ratio at pixel-sized resolution.
The dashed line shows $N_{\text{H$\;$\sc{i}}}^\mathrm{model}/N_{\text{H$\;$\sc{i}}}^\mathrm{thin}=1.3$ and the black solid curve shows cumulative percentage plotted against right-side vertical axis.
Histogram and cumulative percentage of $N_{\text{H$\;$\sc{i}}}^\mathrm{HT}/N_{\text{H$\;$\sc{i}}}^\mathrm{thin}$ ratio of \citet{2003ApJS..145..329H} samples with $|b|>15\arcdeg$ (Table \ref{tab:HTsamples}) are overlaid by the red lines (the former is multiplied by a factor of 250).
\label{fig:model_thin}}
\end{figure*}

\added{
In order to clarify the implications of the histogram of $\int \tau_{\text{H$\;$\sc{i}}}(V)dV$, we plotted the emission-absorption measurements by \citet{2003ApJS..145..329H} in Figure \ref{fig:pix-4arcmin}.
The measurements show that the percentage of the observed $\int \tau_{\text{H$\;$\sc{i}}}(V)dV$ less than 4 is more than 80\%, and that a small fraction of the data points less than 10\% show large integrated \ion{H}{1} optical depth more than 10.
The trend is consistent with the model prediction.
The number of the observed points (see the \cite{2003ApJS..145..329H} data listed in Table \ref{tab:HTsamples}) is, however, very small, $\sim 60$, as compared with the data point of the model, causing larger fluctuations than the model histogram.
}

\added{ 
Table \ref{tab:mass} presents the 3D volume filling factor (Column 9) and the projected covering factor of the CNM (Column 10) for all the four epochs.
The factors vary gently and show no significant difference among at 0.3\,Myrs, 0.5\,Myrs, and 1\,Myr; for these three epochs the volume filling factor ranges from 3.4\% to 5.0\% and the covering factor from 29\% to 36\%.
So, the present small covering factor is not limited to the chosen epoch, 0.5\,Myr, and we consider the model adopted well approximates the general ISM properties in the solar neighborhood.
}

\subsubsection{Observed value of $N_{\text{H$\;$\sc{i}}}$}\label{sssec:obsvaluesNHI}
$N_{\text{H$\;$\sc{i}}}$ does not depend on $T_\mathrm{s}$ while $\tau_{\text{H$\;$\sc{i}}}$ depends on $T_\mathrm{s}$.
$N_{\text{H$\;$\sc{i}}}$ is therefore a more robust measurable quantity than $\tau_{\text{H$\;$\sc{i}}}$.
In order to clarify the impact of the inhomogeneous \ion{H}{1} distribution on the emission-absorption measurements, we used the $N_{\text{H$\;$\sc{i}}}^\mathrm{model}/N_{\text{H$\;$\sc{i}}}^\mathrm{thin}$ distribution in Figure \ref{fig:model_thin_map}.
It is natural that the distribution is qualitatively similar to the CNM distribution.
$N_{\text{H$\;$\sc{i}}}^\mathrm{model}/N_{\text{H$\;$\sc{i}}}^\mathrm{thin}$ has a tight correspondence with $\int \tau_{\text{H$\;$\sc{i}}}^\mathrm{model}(V)dV$; Figure \ref{fig:model_thin}(a) shows $N_{\text{H$\;$\sc{i}}}^\mathrm{model}/N_{\text{H$\;$\sc{i}}}^\mathrm{thin}$ of the present model as a function of $\int \tau_{\text{H$\;$\sc{i}}}^\mathrm{model}(V)dV$.
Due to significant saturation of $W_{\text{H$\;$\sc{i}}}$ in denser regions, $N_{\text{H$\;$\sc{i}}}^\mathrm{model}/N_{\text{H$\;$\sc{i}}}^\mathrm{thin}$ increases with $\int \tau_{\text{H$\;$\sc{i}}}^\mathrm{model}(V)dV$.
$\int \tau_{\text{H$\;$\sc{i}}}^\mathrm{model}(V)dV=4$\,km\,s$^{-1}$ corresponds to $N_{\text{H$\;$\sc{i}}}^\mathrm{model}/N_{\text{H$\;$\sc{i}}}^\mathrm{thin}=1.3$, which we consider as the boundary beyond which a significant underestimate of $N_{\text{H$\;$\sc{i}}}^\mathrm{model}/N_{\text{H$\;$\sc{i}}}^\mathrm{thin}$, more than 1.3 to higher than 2, happens in the optically thin approximation. 

Figure \ref{fig:model_thin}(b) presents a histogram of $N_{\text{H$\;$\sc{i}}}^\mathrm{model}/N_{\text{H$\;$\sc{i}}}^\mathrm{thin}$.
This shows a similar trend with Figure \ref{fig:pix-4arcmin} and indicates that more than 70\% of the pixels have $N_{\text{H$\;$\sc{i}}}^\mathrm{model}/N_{\text{H$\;$\sc{i}}}^\mathrm{thin}$ less than 1.3 and that almost 50\% of them show $N_{\text{H$\;$\sc{i}}}^\mathrm{model}/N_{\text{H$\;$\sc{i}}}^\mathrm{thin}$ less than 1.15; we note that in the present model a ratio of the total \ion{H}{1} mass of the model relative to the optically thin limit is 1.3 (Table \ref{tab:mass}).
It is thus likely that the emission-absorption measurements underestimate the \ion{H}{1} mass for the majority (70\%) of the measurements.

\added{
Figure \ref{fig:model_thin} shows comparisons with the emission-absorption measurements by \citet{2003ApJS..145..329H}, where $N_{\text{H$\;$\sc{i}}}^\mathrm{model}$ is replaced by $N_{\text{H$\;$\sc{i}}}^\mathrm{HT}$ obtained by the authors (Table \ref{tab:HTsamples}).
Figure \ref{fig:model_thin}(a) shows that the behavior of $N_{\text{H$\;$\sc{i}}}^\mathrm{HT}/N_{\text{H$\;$\sc{i}}}^\mathrm{thin}$ is similar to the model prediction, whereas there is a trend that the observed ratio tends to be smaller than the model.
For instance, at velocity integrated \ion{H}{1} optical depth of 4, $N_{\text{H$\;$\sc{i}}}/N_{\text{H$\;$\sc{i}}}^\mathrm{thin}$ is 1.3 in the model and is 1.1--1.2 in the measurements.
This trend becomes more apparent in Figure \ref{fig:model_thin}(b), the histogram of $N_{\text{H$\;$\sc{i}}}/N_{\text{H$\;$\sc{i}}}^\mathrm{thin}$.
The fraction of the $N_{\text{H$\;$\sc{i}}}/N_{\text{H$\;$\sc{i}}}^\mathrm{thin}$ points less than 1.3 occupies more than 90\% as compared with 70\% of the model prediction.
It is not clear what is the cause of the difference.
A possibility may be that the emission-absorption measurements tend to underestimate $N_{\text{H$\;$\sc{i}}}$.
As noted by \citet{2003ApJS..145..329H,2003ApJ...586.1067H}, the fraction of the WNM is still uncertain in the emission-absorption measurements.
Absorption by the CNM also reduces $N_{\text{H$\;$\sc{i}}}$ as shown by the present model (see Table \ref{tab:WNMfraction}).
Correction may be needed to recover the contribution of the WNM in $N_{\text{H$\;$\sc{i}}}$. 
We note that $\int \tau_{\text{H$\;$\sc{i}}}(V)dV$ (Figure \ref{fig:pix-4arcmin}) is not affected by the WNM.
}

In summary, \ion{H}{1} gas has highly complicated sub-pc spatial distribution of $\int \tau_{\text{H$\;$\sc{i}}}(V)dV$ according to the current MHD model.
The fraction of the solid angle subtended by the radio continuum sources is small, less than $10^{-5}$, in the published emission-absorption measurements.
The model predicts a large dispersion in $N_{\text{H$\;$\sc{i}}}^\mathrm{model}/N_{\text{H$\;$\sc{i}}}^\mathrm{thin}$ reflecting the small scale structure of the CNM, which hampers to derive a representative value of $N_{\text{H$\;$\sc{i}}}$ and the emission-absorption measurements are not suited for determining the bulk properties of the \ion{H}{1} gas.

\subsubsection{Possible model dependence}
\begin{figure}
\includegraphics{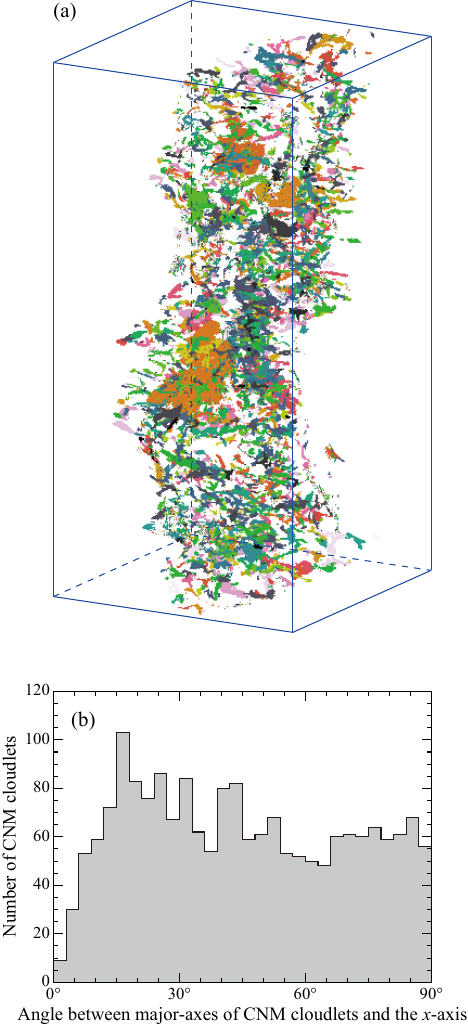}
\caption{
(a) Volume rendering map of the identified CNM cloudlets with $T_\mathrm{s}<300$\,K by using the algorithm described in the Appendix of \citet{2006PASP..118..590R}.
The colors are randomly allocated to the cloudlets.
(b) Histogram of the angle between the elongation of the CNM cloudlets and the $x$-axis.
\label{fig:cloudlets}}
\end{figure}

\added{
In the present paper we used a model of the \ion{H}{1} distribution which was obtained by the state-of-the-art hydrodynamical simulations including the chemical evolution, the magnetic field, and the heating/cooling \citep{2012ApJ...759...35I}.
There are no other simulations at an equal level in the literature; e.g., in \citet{2008A&A...486L..43H} turbulence is included but no magnetic field is incorporated.
The simulations by \citet{2016A&A...587A..76V} only include all these processes, whereas their time step is coarser than the present one, and is not suited to the present purpose to fit the UV measurements of H$_{2}$.
By considering that the model adopted specific initial conditions, it is worthwhile to reexamine if the model is a reasonable realization of the \ion{H}{1} gas in the solar neighborhood and how the initial conditions may affect the present results.
}

\added{
The present simulations assume that the magnetic field is parallel with the \ion{H}{1} flow direction (the $x$-axis).
The directivity of the initial magnetic field, however, does not dominate the gas motion in the 0.5-Myr model adopted, and it is unlikely that the results depend critically on the initial field direction.
This is because of the strong randomization by the density inhomogeneities which causes random deformation of the shock fronts.
In Figure \ref{fig:closeup} we see no strong effects of the initial field direction, except for a slight hint of elongation of the CNM along the $x$-axis.
In addition, we analyzed the CNM in Figure \ref{fig:v-rendering} and identified 9048 CNM cloudlets as shown in Figure \ref{fig:cloudlets}(a) by using the decomposition algorithm described in the Appendix of \citet{2006PASP..118..590R}.
As a result we identified 1880 cloudlets which show significant elongation (see for details Appendix \ref{sec:elongation}), and measured an angle between the elongation of the CNM cloudlets and the $x$-axis.
The histogram of the angle shown in Figure \ref{fig:cloudlets}(b) lends support for the random orientation of the cloudlets.
We also tested several viewing directions in the synthetic observations and confirmed that there is no significant dependence on the directions.
So, we conclude the present results of the synthetic observations are not significantly affected by the initial conditions.
}

\added{
The CNM cloudlets are a natural outcome of the turbulent, thermally bistable ISM.
The present simulations study the shock propagation of the existing two-phase medium that is not thermally unstable.
The shock makes the medium thermally unstable.
Physical thermal conduction is involved in the simulations so that most unstable scale of the thermal instability is resolved.
Even after the saturation of the thermal instability, the CNM cloudlets do not form broad large-scale structure because the CNM cloudlets fragment again due to corrugation instability \citep[see][]{2006ApJ...652.1331I,2008ApJS..178..137S}.
Survival of sub-pc scale cloudlets is confirmed even after $\sim 70$\,Myr evolution in Figure 7 of \citet{2006astro.ph..5528K}.
}

\subsection{The Resolution Issue}\label{ssec:resolutionissue}
The original resolution of our simulations is 0.02\,pc and the synthetic observations were made after smoothing by a factor of two into 0.04\,pc, which was preferred in the present work in order to save the 3-d data size.
The typical width of the CNM filaments is $\sim$0.1\,pc, which is marginally resolved by the present grid 0.04\,pc.
This limited resolution is however not a problem in the above discussion on the covering factor, because the covering factor will not vary significantly as reasoned below; the mass spectrum of the CNM is expressed as $dN/dM \propto M^{-1.7}$ according to the simulations as shown by \citet{2012ApJ...759...35I} and \citet{2008A&A...486L..43H}.
If we assume that the CNM follows this relation down to lower $M$ and that the cross section of a CNM cloudlet $S$ is proportional to \replaced{$M^{2/3}$ if a spherical shape is assumed}{$M$ for a filamentary shape, the covering factor of CNM is then given by $NS\sim M^{0.3}$ that shows small dependence on mass (or scale) of the CNM}.
This means spatial resolution better than sub-pc does not change significantly the covering factor and justifies the present pixel size.
In this connection, we note on the resolution in \citet{2014ApJ...786...64K} used in the two 21-SPONGE papers \citep{2015ApJ...804...89M,2016arXiv161202017M} was 2\,pc for a total length of the simulation box 2 kpc.
This grid size was chosen because these simulations were intended to be applied to a kpc scale \ion{H}{1} distribution.
A 2-pc resolution is however too coarse to resolve the sub-pc filaments of the CNM (see Figure \ref{fig:closeup}) and is not be able to probe details of the emission-absorption measurements discussed in the present paper.

\subsection{Recent Observations of \ion{H}{1} Filaments/Fibers}\label{ssec:recentobs}
The spatial distribution of the CNM has been a subject of \ion{H}{1} observations since 1970s.
Based on aperture synthesis of the \ion{H}{1} absorption toward extended continuum sources, \citet{1973ApJ...184..363G,1973ApJ...184..379G} claimed that the CNM was clumpy.
Subsequent observations in absorption toward extended sources and double continuum sources provided some observational constraints on the CNM distribution \citep{1982ApJS...48..199P,1979ApJ...233..558D}, whereas these studies were not able to constrain the covering factor of the CNM.
\citet{1997ApJ...481..193H} made observations toward sharp gradients in optical depth on scales of milli-arc-seconds and the results support the present conclusion about the compact structure of the CNM. 

It is interesting to note that a few recent papers indicate the existence of CNM filaments similar to those presented in the present work.
\citet{2006ApJ...652.1339M} pointed out such filaments, and \citet{2014ApJ...789...82C} identified the \ion{H}{1} filaments by the Rolling Hough Transformation at 16{\arcmin} resolution.
They named the CNM filaments ``fibers'' which has column density of $5\times 10^{18}$\,cm$^{-2}$, where $N_{\text{H$\;$\sc{i}}}$ of the extended WNM is $10^{20}$\,cm$^{-2}$.
By unsharp masking \citet{2016ApJ...821..117K} identified CNM filaments whose $N_{\text{H$\;$\sc{i}}}$ is $10^{19.1}$\,cm$^{-2}$ in the local ISM.
These CNM filaments/fibers are well aligned with the magnetic field.
\citet{2016ApJ...833...10I}  supported the formation of the CNM filaments and their alignment with the magnetic field based on the MHD simulations.
The filaments/fibers appear to have a small covering factor similar to the present CNM filaments \citep[see Figures 3 and 4 of ][]{2014ApJ...789...82C}.
More quantitatively speaking, we note that the above $N_{\text{H$\;$\sc{i}}}$ of fibers/filaments is crude at best, because the methods based on $W_{\text{H$\;$\sc{i}}}$ alone is not sensitive enough to all $N_{\text{H$\;$\sc{i}}}$ of the CNM; it is difficult to observationally extract $W_{\text{H$\;$\sc{i}}}^\mathrm{CNM}$ from $W_{\text{H$\;$\sc{i}}}$ in Figures \ref{fig:maps}(a) and \ref{fig:maps}(b).
\added{
The column density of the \ion{H}{1} filament/fibers derived by these authors is in the order of $\sim 10^{19}$\,cm$^{-2}$, which  is about an order of magnitude smaller than that of the present CNM clumps $\sim 10^{20}$\,cm$^{-2}$.
Figure \ref{fig:histo_W_N_tau}(b) shows that 60\% of the pixels of the CNM have $N_{\text{H$\;$\sc{i}}}\lesssim 3\times 10^{20}$\,cm$^{-2}$.
Considering that the mass of the CNM is comparable to that of the WNM, it is unlikely that the filaments/fibers of that column density is dominant as the CNM.
In order to test the methods, we applied the unsharp masking to the present $W_{\text{H$\;$\sc{i}}}$ distribution in Figure \ref{fig:maps}(a) and identified filaments.
The result shows that the column density of the filaments is about 20\% of that derived as $N_{\text{H$\;$\sc{i}}}^\mathrm{CNM}$ in the present CNM filaments.
It is possible that the extraction of filaments from $W_{\text{H$\;$\sc{i}}}$ is able to detect only part of the CNM and the \ion{H}{1} column density in \citet{2016ApJ...821..117K} gives a lower value than $N_{\text{H$\;$\sc{i}}}^\mathrm{CNM}$.
The low resolution 16{\arcmin} employed may also be diluting $N_{\text{H$\;$\sc{i}}}^\mathrm{CNM}$ at a sub-pc scale.
Future high resolution \ion{H}{1} studies will help to better quantify $N_{\text{H$\;$\sc{i}}}^\mathrm{CNM}$. 
}

\subsection{Importance of Measuring Accurate $N_{\text{H$\;$\sc{i}}}$; a Potential of the Planck/IRAS Based Method}
A precise measurement of the bulk \ion{H}{1} mass is an important astrophysical issue.
As an example, an application of the \ion{H}{1} measurements is made toward the gamma-ray SNRs where the hadronic process may play a major role in  gamma-ray production \citep[e.g.,][]{2006A&A...449..223A}.
If the hadronic process, basically a proton-proton collision, is working to create gamma-rays in the SNRs, the gamma-ray distribution should resemble the ISM proton distribution for given fairly uniform distribution of cosmic-ray protons in the SNRs.
It is a crucial test to identify the spatial correspondence between the gamma rays and the ISM as already shown in two gamma-ray SNRs, RXJ1713.7$-$3946 and HESSJ1731$-$347 by \citet{2012ApJ...746...82F} and \citet{2014ApJ...788...94F}.

The emission-absorption measurements biased toward a very small volume is not suited to probe the bulk \ion{H}{1} having complicated small-scale fluctuations which acts as the target protons in the hadronic interaction. 
Conversely, the \textit{Planck}/\textit{IRAS}-based method with a larger beam of 5{\arcmin} has a potential as a superior tool for measuring the ISM proton mass and distribution \citep{2014ApJ...796...59F,2015ApJ...798....6F} as demonstrated by a recent work on a gamma ray SNR RXJ0852.0$-$4622 where the \textit{Planck}/\textit{IRAS}-based method is successfully employed to calculate proton density with a high precision in the order of  $\sim$10\% \citep{2017arXiv170807911F}.
The \textit{Planck}/\textit{IRAS}-based method utilizes the data in emission and has an advantage to fully map the \ion{H}{1} gas, which enables to estimate the total \ion{H}{1} mass.
This method is to be better confirmed by proving that the dust optical depth gives a reliable proxy of \ion{H}{1} via comparison with independent ISM measure like $A_{V}$ and gamma-ray counts \citep[e.g.,][]{2013ApJ...763...55R,2016ApJ...833..278M}.
As one of such efforts, in the Perseus cloud, $\int \tau_{\text{H$\;$\sc{i}}}(V)dV$ is estimated for seven radio sources by the emission-absorption measurements \citep{2014ApJ...793..132S} and by the \textit{Planck}/\textit{IRAS}-based method \citep{2017ApJ...838..132O}.
The former gives smaller $\int \tau_{\text{H$\;$\sc{i}}}(V)dV$ from 2\,km\,s$^{-1}$ to 5\,km\,s$^{-1}$, while the latter from 5\,km\,s$^{-1}$ to 16\,km\,s$^{-1}$, with an average ratio of about 2 \citep{2017ApJ...838..132O}.
This difference is not inconsistent with the highly filamentary CNM distribution while the number of sources is limited at present.
Future extension of such a comparison to the other regions by achieving higher sensitivity toward radio compact sources will be important.

\added{
Further, it is required to better quantify the interstellar molecular hydrogen.
$X_\mathrm{CO}$ for converting the CO emission into hydrogen mass is an important factor to estimate the molecular hydrogen and it is generally thought that $X_\mathrm{CO}$ may be uncertain by a factor of 2 in the conventional method \citep[e.g.,][]{2013ARA&A..51..207B}.
The new method to derive $X_\mathrm{CO}$ used by \citet{2014ApJ...796...59F} and \citet{2017ApJ...838..132O} is based on the \textit{Planck}/\textit{IRAS}-data and presented successfully a precise $X_\mathrm{CO}$ distribution in the MBM 53--55 and Perseus clouds.
This method of estimating $X_\mathrm{CO}$ is a promising one to provide $N_\mathrm{H2}$ with a high accuracy of $\lesssim 10$\%, comparable to the present estimate of $N_{\text{H$\;$\sc{i}}}$, indicating a potential of the \textit{Planck}/\textit{IRAS}-based method.
}

\section{CONCLUSIONS}\label{sec:conclusions}
\replaced{}{
In order to gain an insight into the detailed physical conditions of the interstellar \ion{H}{1} gas and their observed properties in the solar neighborhood, we carried out synthetic observations of the interstellar \ion{H}{1} gas at 21\,cm by using the MHD numerical simulations of the realistic inhomogeneous \ion{H}{1} gas which is in the converging flows \citep{2012ApJ...759...35I}.
The simulations incorporate the microscopic processes including the H$_{2}$ formation reactions, the magnetic field, and the heating/cooling.
The simulated \ion{H}{1} gas is highly turbulent and inhomogeneous and is far from equilibrium with a typical dynamical timescale in the order of Myr.
The results were compared with the conventional emission-absorption measurements.
The main conclusions of the present study are summarized as follows;

\begin{enumerate}
\item The present analysis was made for the model at an evolutionary epoch of 0.5 Myrs, which was chosen from the simulation results covering a time span of 10 Myrs.
The model reproduces the distribution of the H$_{2}$ fraction $f_\mathrm{H2}$ which is consistent with the ultraviolet measurements of H$_{2}$.
As shown by the previous works over the last five decades, the \ion{H}{1} gas consists of the two components, the CNM (cold neutral medium) and the WNM (warm neutral medium).
The CNM has spin temperature $T_\mathrm{s}$ ranging from 10\,K to 300\,K and the WNM from 300\,K to 10000\,K.
The density range of the CNM is from 10\,cm$^{-3}$ to $10^{3}$\,cm$^{-3}$ and that of the WNM from 1\,cm$^{-3}$ to 100\,cm$^{-3}$.
The synthetic observations show that the CNM has a small volume filling factor of 3.5\%, whereas the WNM is distributed with a volume filling factor of 96.5\%, while the gas mass of each component is comparable.
These filling factors are consistent with the peak density of the CNM larger than that of the WNM by a factor of $\sim 30$.
The CNM distribution is highly clumpy and filamentary with a sub-pc size scale and the WNM distribution is smooth with much less small-scale structures.
As a result, the CNM covering factor is small, $\sim 30$\%, in the sky.
These represent general properties of the \ion{H}{1} gas in the solar neighborhood. 

\item \ion{H}{1} line profiles were calculated by separating the CNM and WNM.
The CNM is seen as a narrow feature of $\sim 5$\,km\,s$^{-1}$ in half-power full linewidth and the WNM as a wing-like feature spanning over $\sim 40$\,km\,s$^{-1}$.
These properties are consistent with the observed \ion{H}{1} profiles, lending support for the chosen model.
By setting background radiation, absorption line profiles toward radio continuum sources were also synthesized.
The \ion{H}{1} distributions in the sky were reproduced in the 21\,cm line integrated intensity ($W_{\text{H$\;$\sc{i}}}$), the \ion{H}{1} column density ($N_{\text{H$\;$\sc{i}}}$), and the velocity integrated optical depth ($\tau_{\text{H$\;$\sc{i}}}$), both for the CNM and WNM separately.
Saturation of $W_{\text{H$\;$\sc{i}}}$ due to high $\int \tau_{\text{H$\;$\sc{i}}}(V)dV$ greater than 2\,km\,s$^{-1}$ is significant in about half of the \ion{H}{1} gas.
This lends support for the optically thick \ion{H}{1} presented by \citet{2015ApJ...798....6F}.
$\tau_{\text{H$\;$\sc{i}}}$ is dominated by the CNM.
The contribution of the WNM in $\tau_{\text{H$\;$\sc{i}}}$ is negligibly small due to the $T_\mathrm{s}^{-1}$-dependence of the \ion{H}{1} opacity. 

\item The properties of the CNM distribution were compared with the observations in a histogram of $\int \tau_{\text{H$\;$\sc{i}}}(V)dV$.
It is notable that the fraction of $\tau_{\text{H$\;$\sc{i}}}$ less than 4\,km\,s$^{-1}$ corresponds to 80\% of the observed pixels, indicating that the conventional emission-absorption measurements preferentially sample smaller $\tau_{\text{H$\;$\sc{i}}}$ of the WNM.
This reflects the large covering factor of the WNM.
In addition, the nonlinear dependence of $\tau_{\text{H$\;$\sc{i}}}$ as (density)$^2$ causes spatial variation of $\tau_{\text{H$\;$\sc{i}}}$ larger than that of $N_{\text{H$\;$\sc{i}}}$.
The model explains the usual small \ion{H}{1} optical depth obtained by the conventional emission-absorption measurements \citep{2003ApJS..145..329H,2003ApJ...586.1067H}.
Conversely, the fraction of $\int \tau(V)dV$ greater than 2\,km\,s$^{-1}$ is $\sim 50$\%, and the real \ion{H}{1} optical depth is large enough to cause significant saturation in $W_{\text{H$\;$\sc{i}}}$ for about half of the total \ion{H}{1} mass.

\item The present model indicates that $N_{\text{H$\;$\sc{i}}}$ is close to the optically thin limit within a factor of 1.3 at $\sim 70 $\% of the observed pixels and within a factor of 1.15 at $\sim 50$\% of the pixels.
Conversely, pixels with the actual $N_{\text{H$\;$\sc{i}}}$ larger than the optically thin case by a factor of 1.3 occupies $\sim 30$\% of the pixels.
It is usually considered that $N_{\text{H$\;$\sc{i}}}$ is consistent with the optically thin limit, whereas the real \ion{H}{1} mass of the model is 1.3 times larger than the optically thin approximation.
The optically thin approximation thus leads to underestimate the \ion{H}{1} mass by a factor of 1.3, which causes non-negligible errors in estimating interstellar protons. 

\item A detailed comparison of the model with $N_{\text{H$\;$\sc{i}}}$ derived from the conventional emission-absorption measurements indicates that the observed $N_{\text{H$\;$\sc{i}}}$ tends to be systematically smaller than $N_{\text{H$\;$\sc{i}}}$ in the model by a factor of $\sim 1.2$.
It is not entirely clear how $N_{\text{H$\;$\sc{i}}}$ was underestimated in the conventional method.
A possibility may be the uncertainty in the contribution of the WNM whose real intensity is not as acuurate as in $\tau_{\text{H$\;$\sc{i}}}$.
\end{enumerate}

In summary, we studied the detailed properties of the interstellar \ion{H}{1} gas in the solar neighborhood, and made it clear that the CNM has significant sub-pc structures with a small covering factor in the sky.
We showed the observed quantities in the conventional emission-absorption measurements toward radio continuum point sources are subject to an observational bias toward the WNM having a large covering factor.
This bias leads to underestimate $\tau_{\text{H$\;$\sc{i}}}$ and $N_{\text{H$\;$\sc{i}}}$.
Accordingly, the conventional \ion{H}{1} mass is required to be revised upward by a factor of 1.3 in the present model.
The present results provide a step forward toward more accurate determination of the interstellar proton mass.
The mass is crucial for identifying the ISM target protons , for instance, in cosmic-ray proton reactions in the gamma-ray SNRs.
The results are qualitatively consistent with the \textit{Planck}/\textit{IRAS}-based analysis of \ion{H}{1} by \citet{2014ApJ...796...59F,2015ApJ...798....6F}, while a more quantitative pursuit remains as future work, including a test of the nonlinear behavior of the sub-mm dust optical depth due to dust evolution and an extension to the whole sky.
}

\acknowledgments

We are grateful to John Dickey for his thoughtful comments which were valuable in improving significantly the present paper.
The useful comments by the referee helped to improve the content and readability of the paper.
This work was supported by JSPS KAKENHI Grant Number JP15H05694.
Based on observations obtained with \textit{Planck} (\url{http://www.esa.int/Planck}), an ESA science mission with instruments and contributions directly funded by ESA Member States, NASA, and Canada. 
Some of the results in this paper have been derived using the HEALPix \citep{2005ApJ...622..759G} package.
This research has made use of the VizieR catalogue access tool, CDS, Strasbourg, France.



\appendix
\section{Results of the emission-absorption measurements by Heiles \& Troland (2003a)}
Table \ref{tab:HTsamples} gives the results of the emission-absorption measurements by \citet{2003ApJS..145..329H}.
Sixty-one sources which lie at $|b|$ greater than 15{\arcdeg} are selected.
Each column is explained in the footnotes.

\startlongtable
\begin{deluxetable*}{lRRRRRRR}
\tablecolumns{8}
\tablewidth{0pt} 
\tablecaption{Physical parameters of \ion{H}{1} toward radio continuum sources\label{tab:HTsamples}}
\tablehead{
\colhead{Name} & \colhead{$l$} & \colhead{$b$} & \colhead{$W_{\text{H$\;$\sc{i}}}$} & \colhead{$\int \tau_{\text{H$\;$\sc{i}}}^\mathrm{abs}(V)dV$} & \colhead{$N_{\text{H$\;$\sc{i}}}^\mathrm{HT}$} & \colhead{$N_{\text{H$\;$\sc{i}}}^\mathrm{thin}$} & \colhead{$N_{\text{H$\;$\sc{i}}}^\mathrm{HT}/N_{\text{H$\;$\sc{i}}}^\mathrm{thin}$} \\
 & & & \colhead{(K\,km\,s$^{-1}$)} & \colhead{(km\,s$^{-1}$)} & \colhead{($10^{20}$\,cm$^{-2}$)} & \colhead{($10^{20}$\,cm$^{-2}$)} & 
}
\colnumbers
\startdata
    3C18 & 118\fdg 62 & -52\fdg 72 &  283 &  3.26 &  5.99 &  5.17 & 1.16 \\
  3C33-1 & 129\fdg 43 & -49\fdg 34 &  154 &  0.37 &  2.81 &  2.81 & 1.00 \\
    3C33 & 129\fdg 44 & -49\fdg 32 &  154 &  0.27 &  2.78 &  2.82 & 0.98 \\
  3C33-2 & 129\fdg 46 & -49\fdg 27 &  164 &  0.59 &  2.92 &  2.99 & 0.97 \\
    3C64 & 157\fdg 76 & -48\fdg 20 &  333 &  2.02 &  6.33 &  6.08 & 1.04 \\
  3C75-1 & 170\fdg 21 & -44\fdg 91 &  412 &  2.49 &  7.97 &  7.53 & 1.06 \\
    3C75 & 170\fdg 25 & -44\fdg 91 &  409 &  2.73 &  7.89 &  7.46 & 1.06 \\
  3C75-2 & 170\fdg 29 & -44\fdg 91 &  427 &  2.59 &  8.23 &  7.78 & 1.06 \\
    3C78 & 174\fdg 85 & -44\fdg 51 &  497 &  3.97 & 10.06 &  9.08 & 1.11 \\
    3C79 & 164\fdg 14 & -34\fdg 45 &  473 &  3.54 &  9.37 &  8.63 & 1.09 \\
   CTA21 & 166\fdg 63 & -33\fdg 59 &  483 &  2.63 &  9.56 &  8.82 & 1.08 \\
P0320+05 & 176\fdg 98 & -40\fdg 84 &  548 &  5.51 & 11.20 & 10.00 & 1.12 \\
 NRAO140 & 159\fdg 00 & -18\fdg 76 &  603 & 15.96 & 29.49 & 11.00 & 2.68 \\
  3C93.1 & 160\fdg 03 & -15\fdg 91 &  528 &  6.10 & 12.32 &  9.63 & 1.28 \\
P0347+05 & 182\fdg 27 & -35\fdg 73 &  625 &  5.10 & 13.45 & 11.40 & 1.18 \\
  3C98-1 & 179\fdg 85 & -31\fdg 08 &  537 &  3.18 & 10.37 &  9.80 & 1.06 \\
    3C98 & 179\fdg 83 & -31\fdg 04 &  546 &  4.08 & 11.02 &  9.97 & 1.11 \\
  3C98-2 & 179\fdg 82 & -31\fdg 02 &  523 &  2.93 & 10.25 &  9.55 & 1.07 \\
   3C105 & 187\fdg 63 & -33\fdg 60 &  526 & 11.97 & 14.68 &  9.60 & 1.53 \\
   3C109 & 181\fdg 82 & -27\fdg 77 &  767 & 11.73 & 20.82 & 14.00 & 1.49 \\
P0428+20 & 176\fdg 80 & -18\fdg 55 &  970 & 13.57 & 23.89 & 17.69 & 1.35 \\
   3C120 & 190\fdg 37 & -27\fdg 39 &  517 & 10.35 & 15.94 &  9.44 & 1.69 \\
DW0742+1 & 209\fdg 79 &  16\fdg 59 &  134 & -0.26 &  2.43 &  2.45 & 0.99 \\
 3C190.0 & 207\fdg 62 &  21\fdg 84 &  160 & -0.06 &  2.82 &  2.93 & 0.96 \\
   3C192 & 197\fdg 91 &  26\fdg 40 &  216 &  0.50 &  3.97 &  3.95 & 1.01 \\
P0820+22 & 201\fdg 36 &  29\fdg 67 &  231 &  0.47 &  4.23 &  4.22 & 1.00 \\
   3C207 & 212\fdg 96 &  30\fdg 13 &  271 &  2.21 &  5.25 &  4.95 & 1.06 \\
 3C208.0 & 213\fdg 66 &  33\fdg 16 &  165 &  0.27 &  2.99 &  3.02 & 0.99 \\
 3C208.1 & 213\fdg 60 &  33\fdg 58 &  151 &  0.33 &  2.76 &  2.76 & 1.00 \\
   3C223 & 188\fdg 40 &  48\fdg 65 &   57 &  0.27 &  0.96 &  1.04 & 0.92 \\
  3C225a & 219\fdg 86 &  44\fdg 02 &  183 &  0.57 &  3.40 &  3.35 & 1.01 \\
  3C225b & 220\fdg 01 &  44\fdg 00 &  179 &  1.48 &  3.28 &  3.26 & 1.01 \\
 3C228.0 & 220\fdg 83 &  46\fdg 63 &  147 &  0.35 &  2.61 &  2.69 & 0.97 \\
   3C234 & 200\fdg 20 &  52\fdg 70 &   87 &  0.12 &  1.61 &  1.59 & 1.01 \\
   3C236 & 190\fdg 06 &  53\fdg 98 &   64 & -0.14 &  1.21 &  1.18 & 1.03 \\
   3C237 & 232\fdg 11 &  46\fdg 62 &  109 &  0.66 &  2.20 &  1.99 & 1.11 \\
   3C245 & 233\fdg 12 &  56\fdg 30 &  116 &  0.07 &  2.03 &  2.12 & 0.96 \\
P1055+20 & 222\fdg 51 &  63\fdg 13 &   85 &  0.29 &  1.57 &  1.56 & 1.01 \\
P1117+14 & 240\fdg 43 &  65\fdg 78 &   86 &  0.22 &  1.57 &  1.57 & 1.00 \\
 3C263.1 & 228\fdg 27 &  74\fdg 37 &   91 &  0.21 &  1.68 &  1.67 & 1.01 \\
 3C264.0 & 236\fdg 99 &  73\fdg 64 &   95 &  0.25 &  1.73 &  1.75 & 0.99 \\
 3C267.0 & 256\fdg 34 &  70\fdg 11 &  127 &  0.22 &  2.33 &  2.33 & 1.00 \\
 3C272.1 & 280\fdg 63 &  74\fdg 68 &  132 &  0.20 &  2.39 &  2.42 & 0.99 \\
   3C273 & 289\fdg 94 &  64\fdg 35 &  107 &  0.21 &  1.93 &  1.97 & 0.98 \\
 3C274.1 & 269\fdg 87 &  83\fdg 16 &  124 &  0.46 &  2.36 &  2.27 & 1.04 \\
 4C07.32 & 322\fdg 22 &  68\fdg 83 &  113 &  0.49 &  2.11 &  2.08 & 1.02 \\
 4C32.44 &  67\fdg 23 &  81\fdg 04 &   61 &  0.05 &  1.05 &  1.12 & 0.94 \\
   3C286 &  56\fdg 52 &  80\fdg 67 &  110 & -1.26 &  2.04 &  2.02 & 1.01 \\
   3C293 &  54\fdg 60 &  76\fdg 06 &   70 & -0.01 &  1.28 &  1.28 & 1.00 \\
 4C19.44 &   8\fdg 99 &  73\fdg 04 &  144 & -0.39 &  2.65 &  2.63 & 1.01 \\
 4C20.33 &  20\fdg 18 &  66\fdg 83 &  146 &  0.51 &  2.69 &  2.67 & 1.01 \\
   3C310 &  38\fdg 50 &  60\fdg 20 &  190 &  1.57 &  3.71 &  3.48 & 1.07 \\
   3C315 &  39\fdg 36 &  58\fdg 30 &  226 &  2.56 &  4.77 &  4.13 & 1.15 \\
   3C318 &  29\fdg 98 &  54\fdg 78 &  230 &  2.03 &  4.75 &  4.20 & 1.13 \\
   3C333 &  37\fdg 30 &  42\fdg 97 &  247 &  2.31 &  5.09 &  4.51 & 1.13 \\
   3C348 &  22\fdg 97 &  29\fdg 17 &  289 &  2.15 &  5.69 &  5.28 & 1.08 \\
   3C353 &  21\fdg 11 &  19\fdg 87 &  481 &  6.55 & 10.85 &  8.77 & 1.24 \\
 4C13.65 &  39\fdg 31 &  17\fdg 71 &  473 &  2.88 &  9.16 &  8.64 & 1.06 \\
   3C433 &  74\fdg 47 & -17\fdg 69 &  426 &  1.88 &  7.89 &  7.77 & 1.01 \\
 3C454.0 &  88\fdg 10 & -35\fdg 94 &  289 &  0.79 &  5.38 &  5.27 & 1.02 \\
 3C454.3 &  86\fdg 11 & -38\fdg 18 &  349 &  1.74 &  6.53 &  6.37 & 1.03 \\
\enddata
\tablecomments{
Columns (1): name of target, (2) and (3): position in the Galactic coordinates, (4)--(6): \ion{H}{1} parameters given from \citet{2003ApJS..145..329H} dataset; velocity-integrated intensity derived from expected profile, velocity-integrated optical-depth derived from opacity profile, and column density, (7): column density obtained under assumption of optically-thin \ion{H}{1} line, (8): ratio of (6) and (7).}
\end{deluxetable*}

\section{The elongation axes of the CNM cloudlets}\label{sec:elongation}
\added{
The orientation of the elongated CNM cloudlets are determined by using a principal component analysis (PCA).
This method were adopted to determine the position angle of molecular clouds \citep[e.g.,][]{2006ApJ...638..191K,2006PASP..118..590R}.
}

\added{
The density-weighted covariance matrix for a cloudlet is given as
\begin{equation}
C = 
\frac{1}{\sum_{i}n_{i}}
\left(
\begin{array}{ccc}
C_{x, x} & C_{x, y} & C_{x, z} \\
C_{x, y} & C_{y, y} & C_{y, z} \\
C_{x, z} & C_{y, z} & C_{z, z}
\end{array}
\right),
\end{equation}
where
\begin{eqnarray}
C_{x, x} & = & \sum_{i}n_{i}\left(x_{i}-\bar{x} \right)^{2} \nonumber \\
C_{x, y} & = & \sum_{i}n_{i}\left(x_{i}-\bar{x} \right)\left(y_{i}-\bar{y} \right) \nonumber \\
C_{x, z} & = & \sum_{i}n_{i}\left(x_{i}-\bar{x} \right)\left(z_{i}-\bar{z} \right) \nonumber \\
C_{y, y} & = & \sum_{i}n_{i}\left(y_{i}-\bar{y} \right)^{2} \nonumber \\
C_{y, z} & = & \sum_{i}n_{i}\left(y_{i}-\bar{y} \right)\left(z_{i}-\bar{z} \right) \nonumber \\
C_{z, z} & = & \sum_{i}n_{i}\left(z_{i}-\bar{z} \right)^{2}, \nonumber
\end{eqnarray}
$x_i$, $y_i$ and $z_i$ are coordinates in the numerical domain, $n_{i}$ is \ion{H}{1} density of the $i$-th pixel in the cloudlet, and $\bar{x}$, $\bar{y}$ and $\bar{z}$ are the density-weighted mean position of the cloudlet.
We define the elongation axis to lie along the eigenvector with the largest eigenvalue if $|\det C| > 10^{-6}$.
}




\bibliographystyle{aasjournal}
\bibliography{references}


\listofchanges

\end{document}